# Comparing Parallel Functional Array Languages: Programming and Performance

David van Balen[g], Tiziano De Matteis[f], Clemens Grelck[c], Troels Henriksen[a], Aaron W. Hsu[d], Gabriele K. Keller[1], Thomas Koopman[h], Trevor L. McDonell, Cosmin Oancea[a,*], Sven-Bodo Scholz[h], Artjoms Sinkarovs[i], Tom Smeding[g], Phil Trinder[b], Ivo Gabe de Wolff[g], Alexandros Nikolaos Ziogas[e]

[a]*DIKU, University of Copenhagen, Universitetsparken 5, DK-2100, Copenhagen, Denmark*
[b]*School of Computing Science, University of Glasgow, 17 Lilybank Gardens, Glasgow, G12 8QQ, United Kingdom*
[c]*Institute of Informatics, Friedrich Schiller University Jena, Ernst-Abbe-Platz 2, Jena, 07743, Germany*
[d]*Dyalog Limited, Minchens Court, Minchens Lane, Bramley, Hampshire, RG26 5BH, United Kingdom*
[e]*D-ITET, ETH Zurich, Gloriastrasse 35, Zurich, 8092, Switzerland*
[f]*Computer Science Department, Vrije Universiteit Amsterdam, De Boelelaan 1111, Amsterdam, 1081 HV, Netherlands*
[g]*Utrecht University, Heidelberglaan 8, Utrecht, 3584 CS, Netherlands*
[h]*Radboud University, Houtlaan 4, Nijmegen, 6525 XZ, Netherlands*
[i]*University of Southampton, University Road, Southampton, SO17 1BJ, England*

## Abstract

Parallel functional array languages are an emerging class of programming languages that promise to combine low-effort parallel programming with good performance and performance portability. We systematically compare the designs and implementations of five different functional array languages: Accelerate, APL, DaCe, Futhark, and SaC.

We demonstrate the expressiveness of functional array programming by means of four challenging benchmarks, namely N-body simulation, MultiGrid, Quickhull, and Flash Attention. These benchmarks represent a range of application domains and parallel computational models. We argue that *the functional array code is much shorter and more comprehensible than the hand-optimized baseline implementations* because it omits architecture-specific aspects. Instead, the language implementations generate both multicore and GPU executables

---

*Corresponding author

*Email addresses:* d.p.vanbalen@uu.nl (David van Balen), t.de.matteis@vu.nl (Tiziano De Matteis), clemens.grelck@uni-jena.de (Clemens Grelck), athas@sigkill.dk (Troels Henriksen), aaron@dyalog.com (Aaron W. Hsu), g.k.keller@uu.nl (Gabriele K. Keller), thomas.koopman@ru.nl (Thomas Koopman), trevor.mcdonell@gmail.com (Trevor L. McDonell), cosmin.oancea@di.ku.dk (Cosmin Oancea), sven.scholz@ru.nl (Sven-Bodo Scholz), a.sinkarovs@soton.ac.uk (Artjoms Sinkarovs), t.j.smeding@uu.nl (Tom Smeding), Phil.Trinder@Glasgow.ac.uk (Phil Trinder), i.g.dewolff@uu.nl (Ivo Gabe de Wolff), alexandros.ziogas@iis.ee.ethz.ch (Alexandros Nikolaos Ziogas)

from a single source code base. Hence, we further argue that *functional array code could more easily be ported to, and optimized for, new parallel architectures than conventional implementations of numerical kernels.*

We demonstrate this potential by reporting the performance of the five parallel functional array languages on a total of 39 instances of the four benchmarks on both a 32-core AMD EPYC 7313 multicore system and on an NVIDIA A30 GPU. We explore in-depth why each language performs well or not so well on each benchmark and architecture. *We argue that the results demonstrate that mature functional array languages have the potential to deliver performance competitive with the best available conventional techniques.*

## 1. Introduction

Parallel programming is hard. Not only must the developer solve the challenges of sequential software, i.e. specify a correct and efficient algorithm operating over appropriate data structures, but they must also specify parallel coordination, e.g. how the computation is broken into sub-computations, how the sub-computations communicate and synchronize, how they are mapped onto cores, etc. Some parallel hardware is notoriously hard to exploit. For example, obtaining good performance on a GPU requires careful mapping of computations to the hardware, and careful management of the memory hierarchy, e.g., transferring data between GPU and CPU memories or effectively utilizing scratchpad memory as a cache.

The variety of parallel architectures, and their rapid evolution, require performance portability. Parallel language designers aim to generate efficient code for multiple architectures, e.g., enabling a program to be compiled for both multicores and GPUs. If a new generation of hardware increases the number of cores or the amount of memory, the parallel program may need to be refactored for the new architecture. A key objective is to design languages that minimize the refactoring required for a new architecture.

One means of taming the challenges of parallel programming is to focus on some important class of computations. Array languages like APL [1], Fortran [2], or R [3] are designed to express computations on arrays concisely and to implement them efficiently. These languages one way or another generalize operations on scalars to apply transparently to vectors, matrices, and higher-dimensional arrays. Language implementations are carefully designed to exploit the memory hierarchy and often use hardware capabilities such as vector units.

Some array languages are functional, e.g. APL is a design that has stood the test of time. More recently functional array languages like Accelerate [4], DaCe [5], Futhark [6] and SaC [7] have emerged. These languages offer the attractive prospect of low-effort parallel array programming, good performance, and performance portability between architectures.

Although functional array languages adopt a range of design, implementation, and optimization decisions, they have some core commonalities. They use a data-parallel model where bulk parallel operations are performed over the arrays, e.g., mapping a function over every element of an array. This model greatly simplifies parallelization as it avoids explicit synchronization and communication. Parallel functional array languages further avoid typical issues of parallelism like race conditions or deadlocks and lifelocks strictly by design. However, key questions remain. Are such high-level parallel programming models adequately expressive? What are the implications of functional array language design choices? Can the language implementations deliver performance competitive with conventional languages?

We investigate these questions by comparing the designs, implementations, and performance of the five parallel functional array languages Accelerate, APL, DaCe, Futhark, and SaC. We start by introducing their design concepts using the rather simple $N$-body benchmark (Section 2).

We make the following research contributions:

- We systematically compare the designs of five very different functional array languages: Key language design aspects include the typing discipline, the forms of arrays supported, the role of higher-order functions, and the parallel paradigm(s) supported (Section 3).

- We systematically compare the implementations of the five functional array languages. Key implementation aspects are the implementation model (e.g., compiler or DSL), the target architectures, and the supported optimizations (Section 4).

- We show the expressiveness of functional array programming using four challenging benchmarks, namely N-body simulation, MultiGrid, Quickhull and Flash Attention (Sections 6 to 9). These benchmarks represent a range of application domains and parallel computational models and require different optimizations. Rather than choosing simple benchmarks that can easily be expressed and effectively be optimized in all five languages, these challenging benchmarks often take the languages outside of their "comfort zone". Crucially, we argue that *the functional array code is more comprehensible than the hand-optimized baseline implementations*(Section 10).

- We compare the codebase sizes, and hence approximate the programming effort, of the benchmarks in the functional array languages and in the OpenMP and CUDA baselines. We use Source Lines Of Code (SLOC) as a crude but widely accepted measure [8]. A major difference is that the functional array programmer writes a single program that is compiled for multiple targets, where there are separate CPU and GPU baselines. Hence *the total baseline codebase is much larger, at least $10\times$ than any of the functional array codebases. The functional codebases are at least $2\times$ smaller than the CPU baseline, and at least $8\times$ smaller than the GPU baseline codebases* (Section 10).

- We investigate to what extent performant implementations can be generated from a single high-level source for multiple architectures. Can performant code be generated even although the functional array code omits important architecture-specific information, e.g., aspects of the memory hierarchy? We find only partial evidence as only Dace and Futhark consistently achieve good performance on both multicore and GPU (Section 12).

- We report benchmarking results for the five languages on both a 32-core AMD EPYC 7313 multicore system and on an NVIDIA A30 GPU (Sections 6 to 9). We analyze the multicore and the GPU performance of the languages on 39 instances of the four open-source benchmarks to explore why each language performs well, or poorly, on each benchmark and architecture. We find that in 30 % of the instances, at least one language matches or outperforms the baseline; that in 70 % of the instances, at least one language achieves more than 80 % of the baseline performance. In only 6 % of instances, no language achieves more than 50 % of the baseline performance. *We argue that the results show that mature functional array languages have the potential to deliver performance competitive with the best available conventional techniques* (Section 10).

## 2. Parallel Functional Array Languages

To illustrate the five functional array languages compared here, we briefly introduce a simple benchmark, namely (naive) $N$-body simulation. Outlining how $N$-body is implemented introduces the languages and some of the design and optimization choices. More importantly, comparing the implementation in each language with the mathematical specification in the following section, as well as with the CUDA and OpenMP baseline implementations[1], gives a taste of functional array programming.

*2.1. Running Example: the N-body Problem*

The $N$-body problem is a numerical simulation of the gravitational interaction of $N$ particles in space and time, and the naive algorithm is as follows. We are given $N$ particles, each identified by its position $\vec{p_i}$, velocity $\vec{v_i}$ (both 3D vectors), and mass $m_i$ (a scalar). We write

$$\|\vec{p_i} - \vec{p_j}\| = \sqrt{(p_i.x - p_j.x)^2 + (p_i.y - p_j.y)^2 + (p_i.z - p_j.z)^2} \quad (1)$$

for the Euclidean distance between two points $\vec{p_i}, \vec{p_j}$, where $p_i.x$ is the $x$ component of the point $p_i$. We compute the *acceleration* $\vec{a_i}$ of particle $i$ as affected by all other particles with the formula

$$\vec{a_i} = \sum_{j \neq i}^{N-1} \text{force}(\vec{p_i}, \vec{p_j}, \vec{m_j}) \quad (2)$$

---

[1]https://github.com/diku-dk/CFAL-bench/

where
$$\text{force}(\vec{p_i}, \vec{p_j}, \vec{m_j}) = \frac{m_j(\vec{p_j} - \vec{p_i})}{(\|\vec{p_j} - \vec{p_i}\|^2 + \epsilon^2)^{3/2}} \qquad (3)$$
and $\epsilon$ is a softening constant used to avoid excessive interactions between particles that are extremely close. After computing the acceleration, the velocities $v_i$ are updated with
$$\vec{v_i} \leftarrow \vec{v_i} + \vec{a_i} \qquad (4)$$
and once all velocities have been computed, the positions are updated with
$$\vec{p_i} \leftarrow \vec{p_i} + \vec{v_i} \qquad (5)$$
The computation is repeated for some number of steps to simulate the progress of time. As computing the acceleration for a single particle involves looking at all $N$ particles, the total number of interactions computed is $O(N^2)$.

*2.2. The Futhark Language*

Futhark is a statically typed functional array language with support for data parallelism [6]. It is a standalone language with an optimizing ahead-of-time compiler that generates code for a variety of parallel architectures, although its GPU backends are the most developed.

Futhark's syntax and semantics are based on the $\lambda$-calculus. Futhark arrays are modeled as "lists of lists", and the programming experience is very similar to programming in a purely functional subset of a strict language like Standard ML. Futhark supports standard parametric polymorphism and higher-order functions, although with some restrictions: functions are not entirely first class: they cannot be elements of arrays. This is to ensure that higher-order functions can be defunctionalized without requiring control flow or heap allocation of closures [9].

Futhark supports multidimensional arrays of any **rank**, understood as the number of dimensions or axes. However, in contrast to languages such as SaC and APL, Futhark does not support **rank polymorphism**, where functions can be applied to arrays of any rank. This means that to apply an operation across every element of an array, the programmer must explicitly apply map—and perhaps nest it, for a multidimensional array. This makes for verbose code, e.g., scaling a matrix `A` is written as `0.5×A` in APL or as `0.5*A` in SaC, but as `map (map (0.5*)) A` in Futhark. In practice, Futhark programs typically define utility functions for such operations.

In addition to arrays, the supported compound types are records, tuples, and sum types, all of which are structurally typed. Fig. 1 summarizes the syntax through examples. An unusual feature is that array shapes are tracked through a **size-dependent type system**. This detects errors such as multiplying matrices of incompatible size, which is an example of a common source of runtime errors in other array languages, e.g., the dreaded `RANK ERROR` of APL.

Futhark is restricted in various ways to enable efficient execution on parallel architectures with limited resources, such as GPUs. In particular, Futhark does

```
-- Expressions
(a,b,c)                          -- tuples
{a=0, b=1, c=2}                  -- records
[a,b,c]                          -- arrays
loop x = 0 for i < n do x + i    -- sequential loop

-- Types
(i32, f32, u8)                   -- tuple type
{a: i32, b: f32, c: u8}          -- record type
[n][m]i32                        -- array type, of shape n by m
[n]i32 -> [n]i32 -> [n](i32,i32) -- type of zip
```

Figure 1: Some Futhark syntax examples.

not support general recursion, but merely provides tail recursion through a special syntactic construct, named `loop`. The `loop` construct binds the loop parameter (induction variable) x to 0, then evaluates the body n times; binding the result of each evaluation to x, and returning the final value of x, as illustrated in Fig. 1.

Another restriction is that arrays of arrays must be ***regular*** (sometimes also denoted ***rectangular***), meaning all elements of a multidimensional array must have the same shape. This is enforced through the size-dependent type system, with an escape hatch provided via dynamically checked coercions. As an example, the type of the `zip` function, shown in Fig. 1, requires two arrays of the same length.

Parallelism is primarily expressed through ***Second Order Array Combinators*** (SOACs), functions that manipulate an array based on a functional argument. Most SOACs are equivalent to common higher-order functions: `map`, `reduce`, `scan`, `scatter` (parallel write), etc. Parallel code is only generated by the SOACs, and for functions using them. Futhark also allows the programmer to write sequential loops using a `loop` construct, enabling a work/span cost model in the style of Blelloch [10].

The `reduce` and `scan` SOACs require the user to provide an associative operator and neutral element (i.e., a ***monoid***). These properties are not verified by the compiler, providing an invalid operator, like subtraction, may produce unpredictable results.[2] Outside of a few such cases, Futhark is a deterministic language free of any kind of observable side effects.

A more thorough description of Futhark is available from [6].

*Naive N-body simulator.* The Futhark implementation in Fig. 2 illustrates a number of key language design aspects.

---

[2]Addition of floating point numbers is strictly speaking not associative either, yet it is common in Futhark programs. This is often not a problem because passing a nonassociative operator to a reduction does not result in a completely arbitrary *result*, but merely in an arbitrary order of application. The result will still be some sum, although the rounding error will be scheduling-dependent. This is common practice in parallel programming.

```
1   type vec = {x: f64, y: f64, z: f64}
2
3   def vecadd (a: vec) (b: vec) =
4     {x = a.x+b.x, y = a.y+b.y, z = a.z+b.z}
5   def vecsub (a: vec) (b: vec) =
6     {x = a.x-b.x, y = a.y-b.y, z = a.z-b.z}
7   def vecscale (s: f64) (a: vec) =
8     {x = s * a.x, y = s * a.y, z = s * a.z}
9   def dot (a: vec) (b: vec) =
10    a.x*b.x + a.y*b.y + a.z*b.z
11
12  type body = {pos: vec, vel: vec, mass: f64}
13
14  def EPSILON : f64 = 1e-9
15
16  def accel (x: body) (y: body): vec =
17    let r = y.pos `vecsub` x.pos
18    let rsqr = dot r r + EPSILON
19    let inv_dist = 1 / f64.sqrt rsqr
20    let inv_dist3 = inv_dist * inv_dist * inv_dist
21    let s = y.mass * inv_dist3
22    in vecscale s r
23
24  def advance_body (dt: f64) (body: body) (acc: vec): body =
25    body with pos = vecadd body.pos (vecscale dt body.vel)
26         with vel = vecadd body.vel (vecscale dt acc)
27
28  def calc_accels [n] (bodies: [n]body): [n]vec =
29    let move (body: body) =
30      let accels = map (accel body) bodies
31      in reduce_comm vecadd {x=0, y=0, z=0} accels
32    in map move bodies
33
34  def step [n] (dt: f64) (bodies: [n]body): [n]body =
35    map2 (advance_body dt) bodies (calc_accels bodies)
36
37  def nbody [n] (k: i32) (dt: f64) (bodies: [n]body): [n]body =
38    loop bodies' = bodies for i < k do step dt bodies'
```

Figure 2: Futhark $N$-body simulation, representing vectors as records. In a larger application, we would likely define a reusable module for vectors, and perhaps parameterize over the number type. Only the functions `calc_accels`, `step`, and `nbody` involve parallelism, and the latter only by virtue of invoking `step`.

*Vectors.* The three-dimensional vectors `vec` in *N*-body are defined as records with `x,y,z` fields, rather than as three-element arrays. This is different from how many array languages would model vectors, but similar to ML-style languages. Although partially a style choice, there is also an efficiency impact as the Futhark compiler generally assumes that arrays are "large" and sometimes does not manage small arrays efficiently. Hence, it is best practice to model small collections as records. The downside is that utility functions like `vecadd` on line 3 are relatively verbose: the analogous function on array vectors would simply be `map2 (+)`.

*Compound structures.* The `body` type defined on line 12 is a record with three fields, where the `pos` and `vec` fields are themselves records. Other parts of the program use arrays of `body` values. In most languages, such ***Array-of-Structures*** (AoS) are represented by storing the array elements consecutively in memory, which can lead to poor locality. Futhark guarantees that such arrays are instead stored as ***Structures-of-Arrays*** (SoA), i.e., multiple arrays that each contain only primitive elements. In this case, an array of type `[n]body`, like `bodies` on line 28, will be represented as seven arrays of type `[n]f64`. This transformation is invisible to the programmer and improves access locality.

*Custom reduction.* Line 31 performs a reduction with the `vecadd` function using the zero vector as neutral element. The `reduce_comm` function indicates to the compiler that `vecadd` is commutative as well as associative, allowing additional optimization and the generation of faster code.

*2.3. The Accelerate DSL*

Accelerate is a domain-specific language (DSL) deeply embedded in Haskell: an EDSL [4]. It offers the most common Haskell list operations as corresponding parallel operations on multi-dimensional, rank-polymorphic arrays, in addition to permutations, stencil operations, and others. Consider a simple dot-product calculation:

```
import Data.Array.Accelerate
dotp :: (Elt a, Num a)
     => Acc (Vector a)
     -> Acc (Vector a)
     -> Acc (Scalar a)
dotp xs ys = fold (+) 0 (zipWith (*) xs ys)
```

The function is almost identical to a sequential dot-product for two lists in Haskell, but works on one-dimensional parallel arrays (`Vector`). The `zipWith (*)` operation takes the two argument arrays and point-wise multiplies their elements in parallel. The `fold (+)` operation sums the point-wise product of the two arrays.

The operation passed to `fold` must be associative, as it is a parallel reduction. It is worthwhile to have a closer look at the type of `dotp`. The type constraints `Elt a` and `Num a` ensure that the element type `a` is a member of the standard `Num` typeclass, so that numerical operations, such as addition and multiplication are applicable. It also needs to be a member of Accelerate's `Elt` type class that contains all types that may be used as elements of an Accelerate array. This class includes the usual base types, integral types, floating point types, characters, tuples, and more. It can be extended with user-defined types. However, arrays cannot be elements: like Futhark, Accelerate supports only regular (rectangular) arrays, but rather than using a type system, the constraint is expressed by the (exclusive) use of multi-dimensional arrays.

The type constructor `Vector` is a synonym for a one-dimensional array with elements of type `a`, and `Scalar` for a zero-dimensional array. In general, the `Array` type constructor is parameterized by the dimension type, as well as the element type:

```
type Scalar a = Array Z a
type Vector a = Array (Z :. Int) a
type Matrix a = Array (Z :. Int :. Int) a
```

Dimension types are in the type class `Shape`, and are effectively type level lists, with `Z` representing zero dimensions, `(Z :. Int)` one dimension, and so on. For example, a value of shape type `Z :. Int` could be `Z :. 3`.

The arguments of `dotp` are not of type `Vector a`, as one might expect, but instead `Acc (Vector a)`. The reason is that Accelerate is an *embedded* language: the operations do not apply to values, but rather combine abstract syntax trees representing computations. So the type `Acc (Vector a)` represents a *parallel* computation that, when evaluated, returns a value of type `Vector a`. Accelerate offers a second calculation type, `Exp`, for sequential computations.

As an example, the full types of the `fold` and `zipWith` operations are:

```
fold :: (Shape sh, Elt a)
     => (Exp a -> Exp a -> Exp a) -> Exp a
     -> Acc (Array (sh :. Int) a)
     -> Acc (Array sh a)
zipWith :: (Shape sh, Elt a, Elt b, Elt c)
        => (Exp a -> Exp b -> Exp c)
        -> Acc (Array sh a)
        -> Acc (Array sh b)
        -> Acc (Array sh c)
```

As we can see from the functions' types, their first arguments must be sequential (scalar-level) computations. The second argument for `fold`, the source array, is not restricted to be one-dimensional, but can be of any dimensionality $n + 1$. The operation folds over the inner-most dimension, and returns an array of dimensionality $n$. If the array is of dimension $\geq 2$, all inner vectors are reduced simultaneously, using *regular nested parallelism* as those inner arrays all have the same size. These functions, like most of the Accelerate language, are *rank-polymorphic*, and operate on arrays of any dimensionality. So the `dotp` function accepts arrays of any dimensionality greater than 1, and its most general type is:

```
dotp :: (Elt a, Num a)
     => Acc (Array (sh :. Int) a)
     -> Acc (Array (sh :. Int) a)
     -> Acc (Array sh a)
```

The size, or number of elements in, an array is simply an integer, and is not tracked by the type system. As an example, the type system ensures that both

```
 1  data Vec = Vec_ Double Double Double
 2    deriving (Generic, Elt)
 3
 4  type Mass         = Double
 5  type Position     = Vec
 6  type Acceleration = Vec
 7  type Velocity     = Vec
 8  data PointMass    = PointMass_ Position Mass
 9    deriving (Generic, Elt)
10  data Body         = Body_ Position Mass Velocity
11    deriving (Generic, Elt)
12
13  mkPatterns [''Vec, ''PointMass, ''Body]
14
15  instance Prelude.Num (Exp Vec) where
16    (+) = match \(Vec a b c) (Vec x y z) -> Vec (a+x) (b+y) (c+z)
17    (-) = match \(Vec a b c) (Vec x y z) -> Vec (a-x) (b-y) (c-z)
18    (*) = match \(Vec a b c) (Vec x y z) -> Vec (a*x) (b*y) (c*z)
19    negate = match \(Vec a b c) -> Vec (negate a) (negate b) (negate c)
20    abs    = match \(Vec a b c) -> Vec (abs a) (abs b) (abs c)
21    signum = match \(Vec a b c) -> Vec (signum a) (signum b) (signum c)
22    fromInteger i = Vec (fromInteger i) (fromInteger i) (fromInteger i)
23
24  dot :: Exp Vec -> Exp Vec -> Exp Vec
25  dot = match $ \(Vec a b c) (Vec x y z) -> a*x+b*y+c*z
26
27  scale :: Exp Double -> Exp Vec -> Exp Vec
28  scale s = match $ \(Vec a b c) -> Vec (s*a) (s*b) (s*c)
29
30  epsilon :: Exp Double
31  epsilon = constant 1e-9
32
33  pointmass :: Exp Body -> Exp PointMass
34  pointmass = match \case Body p m _ -> PointMass p m
```

Figure 3: Accelerate *N*-body simulation: data type definitions and auxiliary functions.

array arguments to `zipWith` have the same rank, but they can be of different sizes. The result array also has the same rank, and the size is the minimum of the two arrays in each dimension. Values of `Shape` type can be used to specify the size of an array, or to index into an array. For example, the function

```
generate :: (Shape sh, Elt a)
         => Exp sh -> (Exp sh -> Exp a) -> Acc (Array sh a)
```

produces a new array of the size specified by the first argument, and initializes it with the values produced by its second argument, a function from index to value.

*Naive N-body simulator.* The Accelerate implementation in Figs. 3 and 4 demonstrates other characteristics of the language. Fig. 3 defines the data types, and as Accelerate is deeply embedded, user-defined data types are lifted so they can be represented in the abstract syntax tree. This is done automatically for many types by adding `deriving (Generics, Elt)`.

Figure 4 specifies the *N*-body simulation. The `calc_accels` function takes the `bods` vector containing all `n` bodies. To calculate the acceleration of all the combinations of bodies, it creates two temporary matrices by replicating each value `n` times (`bods'`) and by replicating the vector `n` times (`bods''`). The

```
 1  accel :: Exp PointMass -> Exp PointMass -> Exp Velocity
 2  accel = match \(PointMass xpos _) (PointMass ypos ymass) ->
 3    let r = ypos - xpos
 4        rsqr = dot r r + epsilon
 5        inv_dist = constant 1 / sqrt rsqr
 6        inv_dist3 = inv_dist * inv_dist * inv_dist
 7        s = ymass * inv_dist3
 8    in scale s r
 9
10  advance_body :: Exp Double -> Exp Body -> Exp Acceleration -> Exp Body
11  advance_body = match $ \time_step (Body pos mass vel) acc ->
12    let position = pos + scale time_step vel
13        velocity = vel + scale time_step acc
14    in Body position mass velocity
15
16  calc_accels :: Acc (Vector PointMass) -> Acc (Vector Acceleration)
17  calc_accels bods = fold (+) (fromInteger 0) $ zipWith accel bods' bods''
18    where Z_ ::. n = shape bodies
19          bods'  = replicate (Z_ ::. n ::. All_) bods
20          bods'' = replicate (Z_ ::. All_ ::. n) bods
21
22  step :: Exp Double -> Acc (Vector Body) -> Acc (Vector Body)
23  step dt bodies =
24    zipWith (advance_body dt) bodies (calc_accels $ map pointmass bodies)
25
26  nbody :: Acc (Scalar Double) -> Acc (Scalar Int)
27        -> Acc (Scalar Int) -> Acc (Vector Body)
28  nbody dt n k = afst $ awhile (map (< the k) . asnd)
29    (\(T2 x i) -> T2 (step (the dt) x) (map (+1) i))
30    (T2 (gen_input n) (unit $ constant 0))
```

Figure 4: Accelerate *N*-body simulation: main functions.

accel function is then applied on each point-wise pair, and the result is folded into a vector again. Generating the intermediate arrays may seem inefficient, but the implementation eliminates, or fuses, the arrays, and the generated code is a tight parallel loop.

Pattern matching on embedded values is a challenge in an EDSL: if a value has the type Exp a, we do not, in general, have a value of type a available at host language runtime. Instead, we only have an AST that will eventually evaluate to a value of type a. If a is a sum-type, we cannot match in the host language on the actual constructor, but have to generate code for the evaluation of the AST that does the matching. However, to provide a smooth embedding, we want to do this in a way which, for the programmer, looks almost like pattern matching in Haskell.

Accelerate offers a sophisticated workaround for embedded pattern matching [11]. The match function takes a function that pattern matches on a Haskell type, and lifts it into a pattern matching function on the corresponding embedded type. For example, the match on Line 16 in Fig. 3, takes a regular Haskell lambda expression that matches on two values of Vec type, and returns a function of type Exp Vec -> Exp Vec -> Exp Vec. This is specifically needed to support sum types, like Maybe and Either. Data types with a single constructor can be pattern-matched without match, but we still include it in Fig. 3 to show the general case.

All the programmer has to do is add the line `mkPatterns [''Vec, ''PointMass, ''Body]` to the program, which enables this feature for the listed user-defined types. Apart from the use of `match`, the code in Figs. 3 and 4 reads like idiomatic Haskell.

*2.4. The Single Assignment C (SAC) Language*

SaC is a functional array language that aims to provide a platform for easy-to-understand yet generic programs that can be compiled into high-performance parallel codes that run on a variety of different platforms, including multi-core systems, clusters, and GPUs [7].

As the name suggests, SaC can be seen as a side-effect-free variant of C, where assignments are syntactic sugar for nested let-expressions, if-then-else statements are syntactic sugar for conditional expressions, loops are syntactic sugar for tail-recursion, etc. Remarkably, all constructs adopted from C still behave as expected in SaC, and the conceptually very different semantics are surprisingly irrelevant for programming in practice. The key difference between C and SaC that enables this duality between the imperative and the functional world lies in abandoning assignable memory locations. For that reason, SaC does not require any type declaration for variables, but provides type inference instead.

All data in SaC is conceptually immutable and viewed as multi-dimensional arrays. Even *scalars* are considered arrays of rank 0. Arrays are rectangular, and their shape is described by a ***shape vector*** that contains the upper bounds for indices along each dimension (or axis). The length of the shape vector matches the rank of the array. For example, an array of shape `[10,20]` describes a matrix whose first index can range from `0` to `9` and whose second index ranges between `0` and `19`.

Arrays can be nested, but all subarrays must have the same shape (rectilinear arrays). This choice simplifies semantics as nested arrays are indistinguishable from higher-dimensional arrays. For example, `[[1,2,3],[4,5,6]]` may either be regarded as a 2-element vector of 3-element vectors or a rank-2 array of shape `[2,3]`. Heterogeneous nested arrays like `[[1,2,3],[4,5]]` are not permitted.

Any built-in SaC type can be used as an array element type, most C scalar types. In addition, SaC supports stateful types for IO and external types for interfacing with foreign libraries; details can be found in [12]. Furthermore, SaC supports non-recursive records adopting C's `struct` syntax. When used as array elements, records lead to a SoA representation at code generation for improved performance; details can be found in [13].

Key to the expressive power of SaC is the ability to define rank-polymorphic functions in combination with function overloading and rank-polymorphic array comprehensions named ***tensor-comprehensions***. Hence, functions such as `+`, `*`, `==`, etc., are all overloaded by rank-polymorphic versions in the SaC standard library. For example, the subtraction of two arrays of doubles of arbitrary rank and shape is defined as:

```
double[d:shp] - (double[d:shp] arr_a, double[d:shp] arr_b)
{
    return { iv -> arr_a[iv] - arr_b[iv] };
}
```

Here, we overload the subtraction function - for two array arguments `arr_a` and `arr_b`. The types of both arrays are `double[d:shp]`, i.e., arrays of double precision floating point values with the same rank (`d`) and the same shape (`shp`). The result is defined by a tensor comprehension: for each index vector `iv` within the given shape, the result is the difference of the corresponding elements of `arr_a` and `arr_b`.

The equivalence between nested and higher-dimensional arrays naturally leads to generalizations of many standard functions, including selections and operations such as `take` and `drop`. For example, selections with fewer indices than the rank of an array return the corresponding hyperplanes as if the array was nested. For an array of shape `[2,3]`, such as `[[1,2,3],[4,5,6]]`, selecting at the index vector `[0]` results in the first row (`[1,2,3]`), while a selection with the index vector `[0,1]` returns the value `2`.

In addition to tensor comprehensions SaC supports rank-polymorphic reduction on multi-dimensional arrays, such as sum or product of all elements of a numerical array or conjunction and disjunction of the elements of a Boolean array. Reduction is the only second-order construct in SaC; it expects an associative and commutative binary function or operator as argument. As usual, these properties cannot be verified by the compiler for arbitrary user-defined operations. Instead, we consider the use of a function or operator as reduction operation as an implicit assertion of these properties.

Both tensor comprehensions and reductions are syntactic sugar for a more general multi-dimensionsal data-parallel *with-loop* construct. Within the compiler with-loops are uniformly used to represent array operations, optimizations, code generation and parallelization.

The design of SaC as an array language was inspired by APL, whereas the C-like syntax aims at easing code comprehensibility and fostering language adoption by more mainstream programmers. A more thorough description of the design principles of SaC in general and the with-loop in particular can be found in [7].

*Naive N-body simulator.* We define a structure `Body` to capture the position, velocity, and mass for any given body:

```
struct Body {
    struct Vec3 pos;
    struct Vec3 vel;
    double mass;
};
```

The position and velocity vectors are `struct Vec3` records that are defined in the standard library as:

```
struct Vec3 {
    double x;
    double y;
    double z;
};
```

The `acc` function computes the acceleration between two bodies as:

```
struct Vec3 acc (struct Body b, struct Body b2)
{
   dir = b2.pos - b.pos;
   return dir * ( b2.mass / pow3 (EPSILON2 + l2norm (dir)) );
}
```

The definition uses generalized versions of C's built-in functions: subtraction, multiplication, and division are all applied to either arrays of identical shapes, or to a scalar (array with rank 0) argument and a non-scalar array (array with rank greater than 0). Furthermore, we have overloaded these versions not only for all of C's scalar types, but likewise for our record types, here `struct Vec3`.

Finally, we define the time step of the $N$-body algorithm:

```
struct Body[n] timeStep (struct Body[n] bs, double dt)
{
   acc = { [i] -> sum ({ [j] -> acc (bs[i], bs[j]) })}

   bs.pos += bs.vel * dt;
   bs.vel += acc * dt;

   return bs;
}
```

The function takes a vector of `n` bodies `bs` and a time delta `dt`. The first three lines of the body define an array of accelerations for the bodies, `acc`. For each body `bs[i]`, we compute the sum of the accelerations between that body and all other bodies `bs[j]`. As each acceleration value is a record of type `struct Vec3d`, the type of `acc` is `struct Vec3 [n]`. The remaining two lines update the positions and velocities of all bodies, capitalizing on rank-polymorphism to multiply, add, and assign vectors of type `struct Vec3 [n]`.

*2.5. The APL Language*

APL was conceived by Kenneth Iverson in the 1960s [1]. It can be viewed as a dynamically-typed functional language with primitives operating over $N$-dimensional, rectilinear arrays of arbitrary depth (nested arrays) containing character and numeric elements. Arrays may be heterogeneous in both element type and depth, and APL distinguishes the concept of depth, meaning the nesting level of an array, and rank, meaning the number of dimensions in an array.

APL implementations typically support a range of large and small numeric types, converted automatically as needed. This can save space when the representations are small, and usually improves performance, but does come at the cost of checking arrays for their element ranges.

APL's syntax and semantics are unusual, with functions limited to second-order functions of arity less than three (niladic, monadic, or dyadic), written infix. Second-order functions associate to the left, while first-order functions associate to the right. The term "operator" designates second-order functions, and "function" designates first-order functions. Operator applications have higher precedence than function applications. All functions, including primitives, obey the same precedence rules.

APL's main claim to fame is its conciseness, largely achieved by denoting primitive functions and operators as a single symbol. Primitives are defined for all shapes, depths, and sizes of arrays. Exploiting the implicit traversal and iteration patterns in the primitives makes it possible to implement many computations without syntactic control flow.

The following code gives examples of APL's syntax using the Pythagorean theorem, first with parentheses and then using the commute operator to remove the need for them, which is a common programming technique in APL code.

```
h←(+/d*2)*0.5  ⍝ Pythagorean theorem for distance
h←.5*⍨+/×⍨d    ⍝ Same, but using commute and multiply
```

APL's core primitives may be scalar, like the arithmetic functions: these are lifted to apply element-wise for any depth or shape of array. Core functions that are not defined solely over single scalar elements are called mixed functions. The resulting shape or values of these functions depend on a combination of the shapes of the arguments and their values. Among these functions are those to alter or return the shape of an array, reorder its elements, and those to select elements from an array. The following two examples show the function ⊃ for taking the first element of an array, and the function ⍴ for reshaping an array.

```
      ⊃1 2 3   ⍝ Taking the first element of an array
1

      3 4⍴⍳12   ⍝ 3 by 4 Matrix of values [0, 12)
0 1  2  3
4 5  6  7
8 9 10 11
```

In addition to the core functions, APL defines a set of core operators, which define most of the commonly used iteration patterns over arrays. Unlike similar operations in other languages, these operators are defined over all array shapes, allowing for more direct control over the traversal of an array without the need to use more than a single operator. For example, the inner product operator (.) applies over arrays of any rank, including vectors or scalars, as well as matrices, cubes, and arrays with more than 3 dimensions. The inner product is also agnostic to its operands; so, `+.×` implements vector product, matrix multiplication, and batched matrix multiplication in a single function. The expression `∧.=` is a common way to check for equal rows or columns within 2 matrices, but it also tests if 2 simple vectors have equal scalar elements. Likewise, operators like Each (¨) or Reduce (/, ⌿) can apply over any rank or dimension, such as reducing the first or last dimension of an array.

APL permits binding values to alphanumeric names (as well as ∆) using the symbol ← called "gets". For example, `x←1 2 3` binds `x` to the array `1 2 3`. Arrays are indexed with brackets. For example, `A[1;3;0]` denotes the element of `A` at index position(1,3,0), `M[1 2;0 3]` is a submatrix of `M`, and `+⌿[1]` computes the sum (Plus Reduce) along the 2*nd* axis.

Almost all APL primitives have data parallel semantics making them a natural fit for SIMD architectures. There are, however, two challenges: APL does not guarantee the associativity and referential transparency required for the safe parallel execution of operations like scan (`\` or `⍀`); and, the high-level expression of parallel array operations may not necessarily map effectively to a specific SIMD architecture.

*Naive N-body Simulation.* This simple example demonstrates many features of APL as well as some common pitfalls. For a set of `N` bodies in a 3-D space, we have a vector `m` of their masses, a vector `v` of their velocities, and a matrix `p` of their positions. The shape of `p` is `3 N`, which matches the lengths of `m` and `v`. Given two points represented as vectors, the difference between their coordinates can be expressed as `p1-p2`, using the implicit lifting of the `-` function. We can compute the difference of coordinates for all points against all other points with an outer product `∘.-`. For a single dimension `x`, we can compute all differences by writing `p[x;]∘.-p[x;]`, which will give an `N N` matrix of the differences. We can extend this to compute the differences for each dimension in `p` by writing `p∘.-⍤1⊢p`, which will apply `∘.-` over each vector sub-array of `p`, of length `N`, which corresponds to our coordinate dimensions. The result `d` is a `3 N N` cube of the difference of each position against each other position:

```
d←p∘.-⍤1⊢p
```

We compute the distances `h` between each point using the Pythagorean theorem, plus the softening constant `e`:

```
h←(e++/d*2)*0.5
```

Here, the sum `+/` computes over the first axis, which has size `3`; `*` (exponent) is a scalar function that applies to each element in `d`. However, experienced APLers often prefer to avoid needless nesting. We can do so by using `×` instead of `*` to compute squares by using the commute operator `⍨`, which has the following semantics:

```
    X f⍨ Y ↔ Y f X ⍝ Commute arguments
      f⍨ Y ↔ Y f Y ⍝ Replicate right argument in monadic case
```

This allows us to simplify the original expression as follows:

```
h←.5*⍨e++/×⍨d
```

We compute the acceleration of each body by multiplying the mass of the body by the reciprocal of the cube of the distances:

```
a←m×[1]÷3*⍨h
```

Since `h` is an `N N` matrix, the expression `÷3*⍨h` will also be an `N N` matrix, and so we use the axis operator to multiply `m` over the second axis with `×[1]`.

Given accelerations `a`, we can compute new velocities by multiplying the differences in `d` of shape `3 N N` by the accelerations and summing the results for each point, then multiplying by our time step `t`, giving velocity vector `v`:

```
v←t×+/d×⍤2⊢a
```

We use `+/` rather than `+⌿` to multiply over the last axis and not the first, and `×⍤2` to multiply each matrix (rank 2) subarray in `d` by the accelerations in `a`. Using the rank operator `⍤` allows us to map over sub-arrays and compute over each dimension in our 3-D space without repeating the expression or indexing into the array.

We then compute the new positions for each of our points:

```
p←←t×v
```

This single step in the *N*-body computation can then be repeated for `k` steps by writing `step⍣k⊢p v` where `step` is a function that computes the updated values for `p` and `v`.

There is one major flaw in this computation. While the naive *N*-body benchmark assumes the use of the quadratic algorithm, the formulation above may also require quadratic space as it specifies the computation over all points at the same time. So, while the formulation is concise and readily verified, it does not scale to large inputs. A solution is to adapt the code to compute over a single point at a time as follows.

If we alter the shape of the position matrix `p` to be `N 3` rather than `3 N`, we can use `⍤` (Rank) to compute over each point. If we have a function `accel` that computes the accelerations of point `⍺` using all points `⍵`, we can compute the new accelerations for all points with `p accel⍤1 2⊢p`, which applies `accel` for each position in `p` as the left argument `⍺` against all points `p` as the right argument `⍵`.

The definition of `accel` is almost identical to our previous acceleration calculation except that it accounts for the transposed shape of `p` and we no longer need to use the `∘.-` outer product to compute the differences between points:

```
accel←{+⌿d×[0]m×÷3*⍨.5*⍨e++/×⍨d←⍺-[1]⍵}
```

Each time step is computed as above using the following function:

```
step←{p v⊣p←←t×v←v+t×accel⍤1 2⍤⊃p v←⍵}
```

In Figure 5 we combine all these elements to define the *N*-body simulation over `k` steps.

*2.6. The DaCe Framework and SDFG IR*

In contrast to the languages presented so far, DaCe is a framework for mapping code written in high-level programming languages to CPU, GPU, and FPGA programs. DaCe mainly targets high-performance computing (HPC) applications, where two main development roles coexist: the *domain scientist* who provides the mathematical formulations, and the *performance engineer* who optimizes the code for the target hardware. One of DaCe's core principles is the *separation of concerns* of these two roles. The domain scientist provides only a

```
nbody←{k t m←α
  accel←{+⌿d×[0]m×3*⍨÷.5*⍨e++/×⍨d←α-[1]⍵}   ⍝ Acceleration
  step←{p v⊣p+←t×v←v+t×accel⍨1 2⍉⊃p v←⍵}    ⍝ Time step
  step⍣k⊢ω                                   ⍝ k time steps
}
```

Figure 5: APL *N*-body simulation. See text for description of each element.

high-level description of the application's models, unencumbered by hardware-specific code and performance optimizations in general. The performance engineer does not improve the high-level description directly but works on an automatically generated representation amenable to data movement optimizations. Decoupling the application's mathematical description from its optimized implementation streamlines development, making it easy to add new scientific models and to specialize for different hardware architectures.

DaCe's workflow starts with a high-level program written in C [14], Python [5], or another supported programming language. DaCe iterates over the program's abstract syntax tree (AST) to convert the application's high-level description into a graph-based intermediate representation (IR) called Stateful Dataflow multiGraphs (SDFG) [15]. A program's SDFG representation can also be created directly with a graph-based API. In this IR, the program is optimized with DaCe's tools and workflows, either guided by the performance engineer or automatically. Finally, DaCe's backends map the optimized SDFG representation to high-performance C/C++, CUDA/HIP, or HLS codes for CPU, GPU, and FPGA architectures, respectively.

Here we focus on the SDFG IR as it is a parallel functional array language. SDFG is dataflow-based, and its programming model follows three principles:

1. Data containers and computations are separate.
2. Data movement is explicit from data to computations and other data.
3. Control flow is only used to define execution order that is not defined by dataflow.

A detailed description of the SDFG language can be found in DaCe's documentation [16]. Here we describe the IR's core elements as shown in Fig. 6.

SDFG's basic data container is the multidimensional *Array*, a homogeneous random-access data structure. The element types supported include numbers (integers, floats, complex floats) with different bit sizes, boolean values, and strings. An Array's dimensionality (or rank) is constant, but the length of each dimension can be parametric. However, *jagged arrays* are currently not supported. The SDFG does not directly impose constraints on array sizes, but the generated code does, as discussed in Section 4.5. In addition to arrays, the SDFG IR also defines the following data containers:

1. *Scalar*: Memory allocated for a single value; a "0-dimensional Array".
2. *Stream*: Array of First-In-First-Out (FIFO) queues.

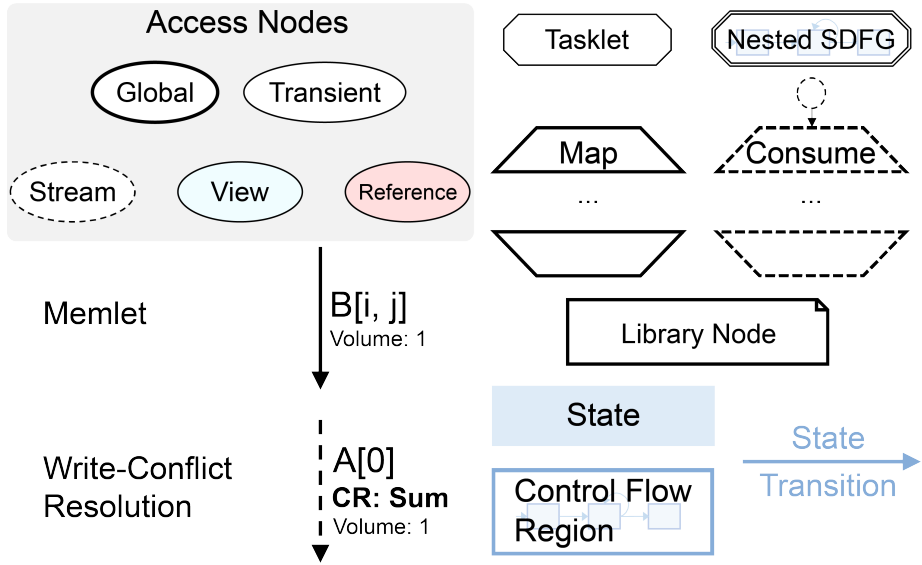

Figure 6: Elements of the SDFG IR, for details see [16].

3. *Structure*: Data containing nested data containers.
4. *View*: Reinterpretation of data, for example, a slice of an Array.
5. *Reference*: A pointer to data containers of the same description (shape, data type, etc.), which may be set to another container dynamically.

Data containers do not appear in the SDFG; rather accesses to data containers are represented by ovals called *Access Nodes*. Global data, i.e., inputs and outputs of an SDFG program, are visualized with a thicker border to distinguish them from transient data defined only in the program's scope. Furthermore, Streams have dotted borders, while Views and References have cyan and red backgrounds, respectively.

Computations are represented by octagons called *Tasklets*, which are typically stateless and operate on scalar values. Tasklets may contain coarser computations, such as C++ code for calling an optimized BLAS library for matrix and vector operations. They can also entirely abstract away another nested SDFG. However, coarse computations are typically expressed through a *Map* or a *Consume*. The SDFG Map represents parametric parallelism as in the other functional array languages. Visually, a Map is described by two trapezoidal nodes, the *MapEntry* and *MapExit*, which pre- and post-dominate its body (an arbitrary subgraph) and are annotated with the Map's iteration space. Consume is the Map's equivalent for streaming computations and is visualized with dotted borders. Another class of computation nodes are *Library Nodes*, which represent high-level operations, such as matrix multiplication or reduction, but do not define their implementation. The latter is freely chosen during DaCe's optimization workflows, allowing Library Nodes to abstract away the use of specialized routines for different target architectures.

The edges connecting data and computations represent data moving from one memory location to another. The SDFG IR annotates its dataflow edges with *Memlets*, which present the data container and the exact subset accessed. For example, in Fig. 6, one of the edges shown reads or writes from data container B the value at index $[i, j]$. Whenever multiple computational units, e.g., threads, concurrently write to the same memory location, a *Write-Conflict Resolution* (WCR) edge must be used. Such cases typically occur inside Maps implementing a parallel fold. WCR edges are dotted to distinguish them from normal dataflow edges, and their Memlet defines how to update the targeted memory location using the current and the new values. For example, the second edge in Fig. 6 writes to index 0 of data container A using the *Sum* method, which atomically adds the new value to the current one.

To represent program execution that does not follow data movement, the SDFG IR encapsulates dataflow in *States*, shown as blue rectangles. States are connected with each other with blue edges annotated with *State Transitions*. These transitions consist of conditions that must be met for the edge to become active and assignments to symbolic variables, such as iteration indices.

The naive $N$-body simulation in Fig. 7 illustrates many of these aspects. The code is "data-centric" Python [5], a subset of Python supported by DaCe and enhanced with elements tailored for developing HPC applications. The equivalent SDFG IR representation is shown on the right. The IR has been partially optimized (Section 4.5) to simplify the correspondence between the Python syntax and the SDFG elements, shown with double-headed arrows.

The program takes three inputs: the bodies' positions `pos` and velocities `vel`, and the timestep's duration `dt`. We include the bodies' masses in the positions data container, i.e., `pos` has four elements for every body. Since the positions and velocities of the bodies are four- and three-element vectors of the same datatype (float64), instead of using structures, we represent them as 2D matrices of size $4 \times N$ and $3 \times N$, respectively. The positions and velocities are overwritten at the end of the timestep's execution and are the program's outputs.

The computation for a single timestep takes place in a single dataflow SDFGState. Map scopes represent operations among arrays. For example, the `dist * dist` product is implemented with a Map scope iterating over the integer space $(d, i, j) \in \{0..2\} \times \{0..N-1\} \times \{0..N-1\}$ and the result is stored in the transient variable `tmp1`. "Reduce" library nodes represent reductions, such as those computing the intermediate data `dist_sq` and `accel`.

## 3. Language Summaries and Comparison

Having outlined the five functional array languages, this section makes a systematic comparison of their features.

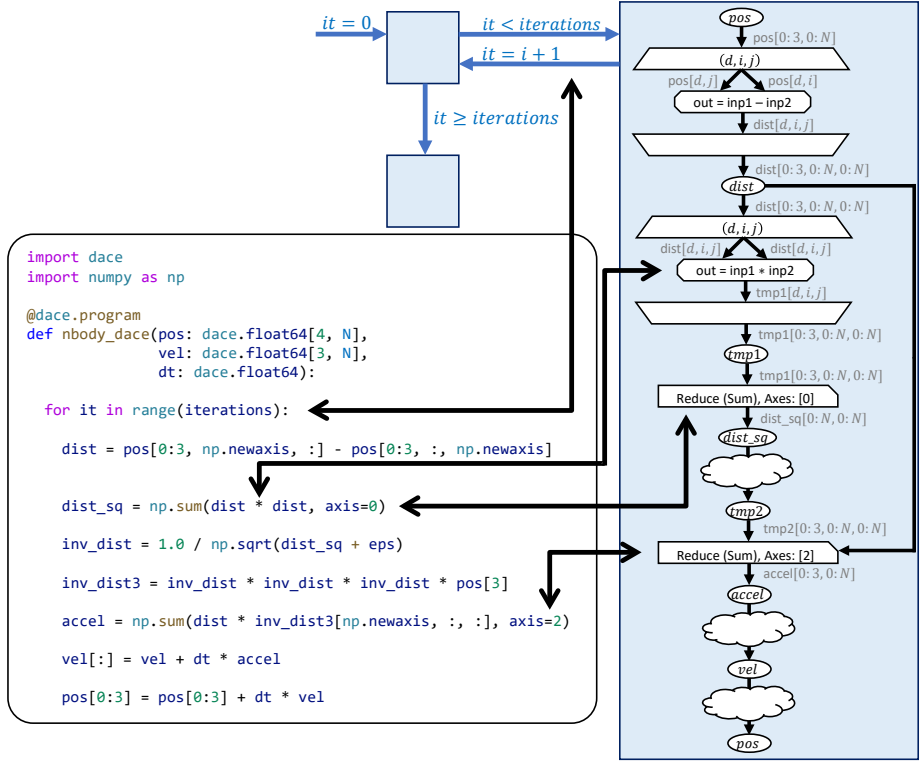

Figure 7: DaCe *N*-body simulation in data-centric Python and its SDFG IR representation. The double-headed arrows match selected computations' Python code with their corresponding SDFG representation. Cloud-like shapes abstract away several SDFG subgraphs containing Map scopes for brevity.

### 3.1. Type Systems

The type systems are one of the most fundamental properties of the languages: Table 1 compares the systems supported by each language. APL is the only dynamically typed language in our comparison, which is consistent with its aim to be expressive. Accelerate and Futhark provide Hindley-Milner-style parametric polymorphism, while SaC is monomorphic, but supports ad-hoc polymorphism through function and operator overloading. In DaCe, any SDFG is monomorphic, with polymorphic frontend languages supported via those languages instantiating any polymorphic constructs at runtime before generating and compiling the SDFG. We discuss implications of the type systems further below.

### 3.2. Array Representation

Array representation is at the core of any array language, and Table 2 compares the representations supported by the languages. APL is the only language in our comparison that supports jagged arrays, i.e., multidimensional arrays where inner arrays can have different shapes. All other languages support only rectilinear arrays. APL represents jagged arrays via nesting, where each dimension of jagged size introduces another level of nesting. APL has a primitive concept of uniform and non-uniform depth, indicating the level of array nesting. In the case of ragged arrays, the depth indicates the number of jagged dimensions, while the shape at each depth of the array thus gives the dimensions of the uniform dimensions of the shape up to the next jagged hyperplane of the array. In the case of a matrix with jagged columns, the depth 0 array would be a vector whose length is the number of rows in the matrix while each element of the vector would be another vector whose length is the number of columns in that row. As such, it can be argued that APL also supports only rectilinear arrays, but it does provide a natural (and commonly used) mechanism for representing jagged arrays.

Table 1: Type Systems in the Functional Array Languages

|  | Accelerate | APL | DaCe | Futhark | SaC |
|---|---|---|---|---|---|
| Discipline | Static | Dynamic | Static | Static | Gradual |
| Parametric polymorphism | ✓ | ✓ | Inherited | ✓ | ✗ |
| Dependent | ✗ | ✗ | ✗ | ✓ | ✓ |

Table 2: Array Representation in the Functional Array Languages

|  | Accelerate | APL | DaCe | Futhark | SaC |
|---|---|---|---|---|---|
| Shape | Rect. | Rect. | Rect. | Rect. | Rect. |
| Rank Polymorphism | ✓ | ✓ | ✗ | ✗ | ✓ |
| Jagged Arrays | ✗ | ✓ | ✗ | ✗ | ✗ |
| Heterogeneous | ✗ | ✓ | ✗ | ✗ | ✗ |
| User-defined types | ✓ | — | ✓ | ✓ | ✓ |

While the flexibility of APL is widely used by its programmers, we shall see later that high-performance APL programs tend to use a subset of features similar to those supported by the other, statically-typed, languages. In particular, jagged arrays are used sparingly.

Rank polymorphism, where functions can be applied to arguments of any rank, is a cornerstone of classic array programming, and is supported in Accelerate, APL, and SaC. It arises naturally from how these languages view the shape of an array as distinct from its elements. Specifically, APL, SaC, and Accelerate view an array as a flat *value vector* of scalars, along with a *shape vector* of integers, where the product of the shape vector equals the length of the value vector. Data parallel operations, such as for example addition, thus operate directly on the elements of the value vector, and simply propagate the shape vector. This enables an efficient value representation and ease of implementation, no matter the rank of the arrays.

Table 3: Parallel Computation Paradigms in the Functional Array Languages

|  | Accelerate | APL | DaCe | Futhark | SaC |
|---|---|---|---|---|---|
| **Data parallelism** | Explicit | Implicit | Explicit / Implicit | Explicit | Implicit |
| **Higher-order functions** | Restricted | ✓ | Frontend-dependent | Restricted | ✗ |
| **Recursion** | ✗ | ✓ | ✗ | ✗ | ✓ |
| **Nested Parallelism** | ✗ | ✓ | ✓ | ✓ | ✓ |
| **Task parallelism** | ✗ | Fork-Join | Explicit | ✗ | ✗ |
| **Determinism** | ✓ | ✓ | Code dependent | ✓ | ✓ |

Accelerate, APL and SaC even model scalars as 0-dimensional arrays, containing a single-element value vector, meaning that operations on arrays are not a special case. Accelerate distinguishes arrays and scalars in the type system, mostly to eliminate arrays-of-arrays. SaC has different runtime representations of arrays depending on static rank and shape knowledge. Therefore, SaC efficiently stores 0-dimensional arrays on the stack, instead of allocating and managing heap space, in the usual case.

Futhark is an outlier: its array model is more inductive and views multi-dimensional arrays simply as arrays-of-arrays. Futhark then uses a size-dependent type system to avoid jagged arrays. At run-time, the elements of a Futhark array value will still be a single flat vector of values, similar to the rank-polymorphic languages.

*3.3. Parallel Computation Paradigms*

Table 3 compares the parallel computation paradigms supported by our five languages. All of the languages provide bulk *data parallelism*, i.e., parallel operations on entire arrays, but they do so in different ways. Accelerate and Futhark rely on *parallel combinators*, or higher-order functions, which must be explicitly invoked by the programmer. In APL, much parallelism is *implicit* in rank-polymorphic array operations. In SaC, parallelism ultimately arises from the use of tensor comprehensions and reductions, which in many cases are located within rank-polymorphic library functions. Compiler and runtime system jointly decide which concurrent operations are actually run in parallel and how) on the target architecture and which not. In DaCe, programmers can *explicitly* specify the parallelism by creating an SDFG containing parallel operations like Map. However, if the SDFG is generated by a frontend for a language with array operations, such as Python with NumPy arrays, the parallelism can be expressed in the form exposed by the frontend.

*Higher-order functions.* All of our five languages restrict the use of *higher-order functions* or entirely disallow them. APL only supports second-order functions, meaning that higher-order functions must take only first-order functions or arrays as arguments and can only return first order functions. APL's notation, outside of implementation-specific extensions, does not support the creation of closures in the form of first-class procedures.

Futhark supports higher-order functions, but they are not fully first-class. In particular, they may not be returned from conditional expressions, be array elements, or be carried across sequential loops. These restrictions ensure that higher-order constructs can be completely defunctionalized without any runtime overhead [9].

In Accelerate, higher-order functions can only be parameterized with sequential functions. Effectively, this results in restrictions similar to Futhark.

SaC does not support higher-order functions in general, but features a language construct that implements reductions using an associative and commutative reduction operation. Programmers are responsible to ascertain these properties in the case of user-defined operations.

DaCe supports a limited set of higher-order functions. In particular, depending on the utilized frontend, higher-order functions can be defunctionalized or otherwise evaluated prior to the generation of the SDFG, very similar to Accelerate. Further, the Python frontend also includes custom reductions where the operation is defined via a lambda function, method parameters as callbacks to the CPython interpreter, and advanced indexing of NumPy arrays, which filters the underlying data based on another array.

With the exception of APL, these limitations are all rooted in implementation concerns: in high-performance code, function values are somewhat awkward, as they require application through function pointers. Further, arrays of function values can be difficult to represent efficiently, as even when their type is the same, the actual function (and closure environment) may differ widely. The use of higher-order functions in Futhark, Accelerate, and DaCe is specifically restricted to ensure that they can be compiled away (*defunctionalized*) early in the compilation process, thus avoiding these representation challenges.

*Nested parallelism.* A key property is whether the parallel constructs can be *nested*. Nesting constructs increases expressive power and aids performance, e.g., to specify how computations are mapped to GPU thread blocks, warps/wavefronts within a block, and threads within a warp. Futhark and DaCe allow nested parallelism, although in the case of Futhark's GPU backend, the nesting must be *regular*: the sizes of inner parallel loops must be invariant to outer loops. This implementation restriction does not apply when using the multicore CPU backend. Although Accelerate does not allow parallel functions as arguments to higher-order parallel functions, it supports regular nested parallelism via rank-polymorphism. For example, a fold operation applied to a multi-dimensional array will fold, in parallel, over all arrays in the innermost dimension. The languages with implicit parallelism, APL and SaC, implicitly allow for nested parallelism by using operations that give rise to parallelism in a nested fashion.

Of the languages studied, only APL and DaCe also provide *task parallelism*. However, this paper focuses on data parallelism, and we do not explore the combination of data and task parallelism any further.

*Determinism.* Some parallel functional languages provide *deterministic parallelism* where the value computed by the program is always the same regardless of how it is executed, and in particular, is the same as the value computed by a sequential execution. Parallelism in imperative languages is often non-deterministic, due to factors such as random scheduling. Non-determinism is recognized as one of the greatest challenges for parallel programming as it makes understanding, debugging, testing, and securing a program hard [17]. Determinism avoids this entire class of problems, and in particular guarantees the absence of race conditions.

Accelerate, Futhark, SaC and the data parallel fragment of APL that we consider here, guarantee determinism in principle, although with caveats. Specifically, the languages provide parallel reductions that are only deterministic if the provided operator is associative, a property that usually cannot be verified for user-defined operations. Moreover, all five languages unanimously consider floating point addition and multiplication to be associative, whereas they are strictly speaking not. This is a common feature across parallel computing in general because reductions with floating point addition and multiplication are common in numerical codes and floating point arithmetic is only an approximation of real arithmetic. However, the precise binary representation of a reduction may depend on runtime parameters beyond the control of programmers. This source of non-determinism is widely accepted in parallel computing.

A DaCe program may be deterministic: the generated code for a given SDFG is always exactly the same, but the program may inherit non-determinism from the backend languages, e.g., from CUDA or C/C++ with OpenMP.

## 4. Language Implementations and Comparison

This section outlines the implementation of the five functional array languages, before making a systematic comparison of languages covering implementation model, execution targets, and key optimizations supported.

*4.1. Implementing the Futhark Language*

The Futhark compiler is a heavily optimizing, but architecturally conventional, ahead-of-time compiler. It generates code for both GPUs (targeting OpenCL, CUDA, or HIP) or CPUs (targeting POSIX Threads). Each compilation target has a distinct optimization pipeline, although with significant overlap in passes. Compilation begins by compiling away module-level constructs, monomorphizing polymorphic functions, defunctionalizing higher-order functions, and flattening compound types, such as tuples. Arrays of compound types are represented as *multiple* arrays, each storing only primitive scalar types. For example, a source array of type `[n](i32,bool)` is represented in the compiler as two arrays of type `[n]i32` and `[n]bool`. This is generally known as a structure-of-arrays (SoA) representation. Hence the remainder of the compilation operates on a first-order, monomorphic program with only primitive types and arrays of primitives, and a handful of built-in second-order array combinators (SOACs) such as maps, scans (prefix sum), reductions, and generalized histograms [18].

The compiler performs standard optimizations, such as copy propagation, constant folding, inlining, common subexpression elimination, loop hoisting, and so on. The most important early optimization is **fusion** (or *deforestation*) [19], where adjacent array traversals are combined to avoid materializing intermediate arrays. Many fusion rules are used, and a simple example fuses `map`s, i.e. replaces traversals for `f` and for `g` with a single traversal for the composition `f o g`:

$$\texttt{map f} \circ \texttt{map g} \Rightarrow \texttt{map (f} \circ \texttt{g)}$$

Futhark supports nested parallelism, but some compilation targets, like GPUs, only efficiently support a fixed number of parallel levels, and there are significant performance implications in how they are used. Therefore, one of the most important transformations is *flattening* [20], where multiple levels of application-level parallelism are collapsed and assigned to different hardware levels. This is achieved by **incremental flattening** [21], which uses map fission and loop interchange to create semantically equivalent code versions that utilize more and more levels of application parallelism. Essentially, using a top-down program traversal, whenever a new `map f` operation is discovered, the analysis:

(1) creates a first version corresponding to a CUDA kernel in which each thread (independently) executes an application of $f$.

(2) creates a second version, dubbed *intra-block kernel*, in which the `map` parallelism discovered so far is mapped on the CUDA grid, and the parallelism inside $f$ is flattened and mapped to the CUDA block level; this has the benefit that intermediate arrays are created and reused from fast memory.

(3) continues flattening recursively (by means of map fission and loop interchange) in a parallel context that is extended with the current map.

The resulting code versions are combined together into a program that branches depending on some dynamic program measure with a threshold. The best combination of code versions is derived by autotuning the threshold values [22].

Subsequent optimization passes (i) rewrite code using accumulators in terms of more specialized constructs such as reduce(-by-index) [23], (ii) optimize temporal locality, e.g., tiling in registers and shared memory is performed when an array that is invariant to one of the parallel dimensions is "streamed" through an operator, and (iii) minimize the footprint of scratchpad memory and eliminating unnecessary copy operations [24].

Finally, SOACs benefit from specialized GPU code generation: e.g., the implementation of scan [25, 26] uses Merrill and Garland's single-pass algorithm [27], and that of reduce-by-index uses a combination of multi-pass and multi-histogram techniques to improve locality and reduce atomic conflicts [18].

*4.2. Implementing the Accelerate DSL*

As Accelerate is a deeply embedded language (Section 2.3) constructs do not directly operate on arrays or scalar values. Instead, constructs operate on abstract syntax trees (ASTs) that represent computations of arrays and scalars. Users are mostly shielded from this fact, and write programs that look like regular Haskell programs. An embedded implementation brings a number of advantages: much of the mature host language infrastructure, such as the GHC compiler front-end, run-time system, and some of the memory management can be used. However, embedding also means that generating the code incurs runtime overhead. The implementation aims to minimize the overhead by optimizing compilation time and caching compiled code so that if the same code is run multiple times, it is only compiled once.

Accelerate offers currently three main, fully implemented back-ends: one that targets NVidia GPUs generating PTX code via LLVM, a multicore CPU back-end (also via LLVM), and a sequential interpreter for debugging. However, most of the implementation is platform-independent.

As for Futhark, and many other languages where array operations are expressed as higher-order functions, fusion is a core optimization [4]. Similarly, regular nested operations are flattened, and the necessary monomorphic instances of polymorphic operations generated. Scans and folds in inner loops are only executed in parallel if the range of the parallel outer loop is too small to generate sufficient parallelism. The mapping of convenient surface types (which the user programs with), such as arrays of compound types, to machine-friendly arrays of primitive types is also similar to Futhark and other array languages in that it results in a SoA representation. In Accelerate, the transformation to SoA form is implemented as part of the translation from surface to internal types, and exploits GHC's type families and associated types [28, 29].

In addition to language optimizations, Accelerate requires some additional optimization specific to embedded languages. One such optimization is *sharing recovery* that maps user-friendly higher-order embedding with lambda abstractions and let-bindings, into a first-order embedding [4]. Runtime compilation also allows for additional specializations of the code.

*4.3. Implementing the Single Assignment C Language*

Although the name and syntax of Single Assignment C may suggest a lightweight implementation on top of a C compiler, the reality is very different. The sac2c SaC compiler is a many-pass compiler for a language that is syntactically very similar, and often identical, to C; Fig. 8 provides an overview.

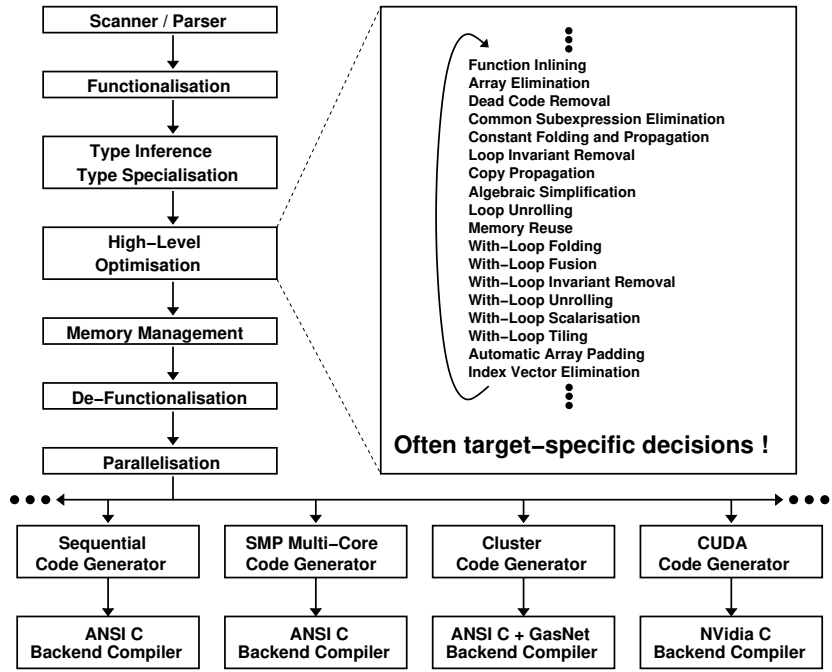

Figure 8: Structural organization of the SaC compiler sac2c

The first block of code transformations is ***functionalization*** where the intermediate code is transformed from the imperative-flavored source SaC into a representation that makes the underlying functional semantics more explicit. For example, loops are transformed into tail-recursive functions, if-then-else constructs into conditional expressions, etc.

The next block of compiler passes performs type inference and checking. In contrast to C, variables in SaC can be introduced on the fly and without declaring their types.

The high-level optimization pass is crucial, and most optimizations are architecture-independent. Figure 8 also enumerates the most important optimizations. The optimizations include both conventional compiler optimizations, like copy propagation, and specific optimizations for functional array processing, like the various with-loop optimizations listed in Fig. 8. The SaC compiler sac2c aims at fusing with-loops in three dimensions: Vertically, it fuses with-loops where the result of one becomes the input to another; we call this with-loop-folding. Horizontally, it fuses (or tuples) with-loops that compute results based on the same or at least overlapping set of argument arrays; we call this with-loop-fusion. And in the remaining dimension, we fuse nested with-loops where one computes the element-wise value of enother; we call this with-loop-scalarization. With-loops are SaC's one-stop-shop solution to both specify and internally represent data-parallel array operations; for details see [30]. Any language-level array operation is converted into some with-loop, including the tensor comprehensions introduced in Section 2.4. All optimizations benefit from the underlying purely functional semantics of SaC, and can usually be applied more aggressively than with an imperative interpretation of syntactically identical or very similar C code.

The memory management pass associates abstract arrays with concrete memory. This is where we control memory reuse and go to great lengths to counteract the aggregate update problem, both key aspects to achieving high performance with functional arrays. Details can be found in [31].

The next pass *defunctionalizes* the IR for the imperative target architectures. For example, tail-recursive functions are converted back into loops, etc.

Parallelization is a surprisingly small pass, largely because the functional array code is typically highly parallel in nature, and all array operations are internally represented by a single IR construct: the with-loop. With-loops typically have more concurrency than most architectures have execution units. For this reason, the compiler focuses on how much work to map to each processor (be it a core, a node, or a thread).

In a final pass, the SaC compiler selects and parameterizes the backend compiler to produce a linked executable. A more detailed description of the SaC compilation pipeline is available in [32].

*4.4. Implementing the APL Language*

APL has multiple implementations, and in the benchmark sections we use the Dyalog interpreter on the multicores and the Co-dfns compiler on the GPUS. Dyalog is a bytecode interpreter for APL source token streams implemented in C/C++. The interpreter calls a specialized function for each APL primitive. Most primitives have multiple implementations designed to run more effectively on different input shapes, data ranges, etc. For some primitives, a minor degree of multi-threading is used if the data ranges are large enough, but this appears to have little impact on the benchmarks used here.

The design and architecture of the Co-dfns compiler is novel [33], and based on an extension of the NanoPass architecture [34]. It consists entirely of small, incremental, data-parallel passes written in APL to construct the entire compiler pipeline. It uses almost zero abstractions over pure APL arrays, where the AST is represented using an SoA design and is exceptionally efficient in its data representation. Outside of the major pipeline functions encapsulating the parser, compilation, and code generation phases, it contains almost no function abstraction nor many internal temporary variables, but only primitive APL function expressions over a few AST fields and temporary variables. The code, thus, lacks explicit looping constructs, branching, conditionals, or other explicit control flow outside of the use of the power operator to create iteration to fixed points for certain passes. As a result, the compiler is exceptionally compact, while being data-parallel by construction and thus GPU compatible.

The Co-dfns compiler optimizes very little at compile time. Efficient data representations, fusion of APL primitive calls, GPU acceleration, inlining of functions, etc., all utilize runtime implementations or the backend compiler, such as a C compiler. Fusing primitive calls in order to reduce the materialization of intermediate arrays and to reduce the number of GPU kernels generated is accomplished using existing JIT and GPU libraries. [35] This simplifies the compiler while at the same time enables features such as fusion to occur across user-defined function call boundaries. However, the computations being fused must be combinations of pre-existing APL primitives. For example, reductions using max, min, or plus can all be fused with scalar primitives in a single pipeline, but reduction with a user-defined function cannot be fused. The runtime can detect the use of some known functions, and if idiomatic combinations are used, such as $+.\times$ for matrix multiplication, the runtime dispatches to an optimized library function specialized based on a combination of the datatypes involved and the functions being called. However, these specializations are limited to what the runtime can see. Thus, optimizations like algebraic-style expression simplification are not applied.

The latest version of the Co-dfns compiler uses a runtime implemented in APL, with a small subset of functionality provided by the underlying host platform, such as C or JavaScript, in the form of a small kernel library. This is achieved using a language-agnostic foreign function interface.

*4.5. Implementing the DaCe Framework*

DaCe expresses program optimizations as graph transformations on the SDFG IR [15]. Transformations are applied in optimization passes, and may alter any aspect of the graph representation. They may be applied to the whole program, or only to specified subgraphs. Multiple passes can be combined into *pipelines*, ensuring that pass dependencies are met. DaCe provides a large library of transformations, passes, and pipelines. The framework includes user-level APIs to allow the development of new transformations and pipelines [36, 37]. Although many of the built-in passes and pipelines can be applied automatically, the user can opt to optimize a DaCe program manually. Manual optimization can use a set of APIs or visual tools , that guide minimizing data movement [38]. DaCe also provides APIs for instrumenting, optimizing, and auto-tuning SDFG programs through performance modeling with built-in or external tools [39, 40].

We illustrate some built-in data-centric transformations for the *N*-body program, focusing on the single iteration of the algorithm shown in Fig. 9a. The first optimization expands the Reduce Library Nodes to Map scopes, and must introduce Write Conflict Resolution (WCR) on the output edges, as multiple Map iterations write to the same memory location (Fig. 9b). Next, we permute the dimensions of selected Map scopes and apply the *MapExpansion* transformation so that all subgraphs have an outer iteration over the bodies using index *i* (Fig. 9c). Finally, we apply the *SubgraphFusion* transformation to fuse all subgraphs up to (and including) the computation of the updated velocities (vel) under a common outer Map scope (Fig. 9d). The transformation does not apply to the Map scope that updates the bodies' position, as that would produce read/write conflicts on pos. Fusing the Map scopes renders all intermediate results thread-local and reduces their dimensionality. For example, dist and accel become three-element vectors, and inv_dist3 reduces to a scalar value.

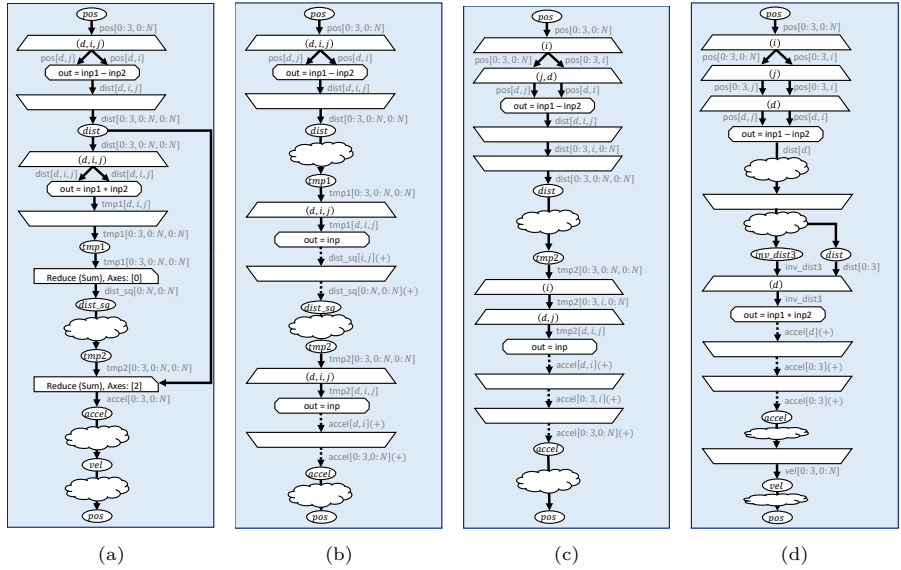

Figure 9: Optimization workflow for the acceleration computation of the *N*-body program in DaCe: (a) unoptimized SDFG IR, (b) Library Nodes' expansion, (c) Map scope dimensions' permutation and *MapExpansion*, (d) *SubgraphFusion*.

Specialization for a specific architecture like a GPU also occurs in the SDFG IR. Data containers have storage types, and Maps have schedule types (Section 2.6). For example, to "convert" the *N*-body program for execution on a GPU, we simply switch the program's inputs and outputs to *GPU-global* memory. Similarly, the Maps' schedules must be changed to *GPU-device* and *GPU-thread-block*. Storage and schedule changes can be applied manually (programmatically or using visual tools) or automatically by applying device-specific transformations, such as *GPUTransform*, which iterates over the SDFG and updates the schedule of Map scopes deemed suitable for GPU execution. Data movement between the host and device is explicit in the SDFG and can also be automated with the same device-specific transformations.

Since a program's SDFG representation includes scheduling and storage specification, DaCe's code generation is relatively simple. The first step is to traverse the SDFG to determine the lifetime of all transient variables, mark-ing the appropriate states for allocation and de-allocation. Subsequently, DaCe partitions the SDFG into subgraphs defining control-flow scopes. For loops and structured control, DaCe generates the corresponding C/C++ syntax, e.g., `for`, `while`, `if`, etc. Unstructured control flow is generated as a state machine with `goto`s. The body of the control flow scopes is generated state-by-state in a topological order. Each state is code-generated in three steps: (1) transient allocation, (2) dataflow code generation, and (3) transient de-allocation. To generate code for dataflow, DaCe traverses the graph in topological order, generating appropriate code based on the schedule and storage types, e.g., a map with a *CPU-multicore* schedule generates an OpenMP for-loop, where a GPU-related schedule generates a CUDA/HIP GPU kernel.

*4.6. Language Implementation Comparison*

This section compares the implementations of the five functional array languages. There are a variety of implementation techniques: the Futhark and SaC compilers, the Accelerate DSL, the DaCe framework, and the APL interpreter/-compiler. The key differences are summarized in Table 4.

Table 4: Table summarizing key differences between language implementations.

|  | Accelerate | APL | Futhark | DaCe | SaC |
|---|---|---|---|---|---|
| **Implementation model** | | | | | |
| Standalone | ✗ | ✓ | ✓ | ✗ | ✓ |
| Implementation language | Haskell | APL | Haskell | C/Python | C |
| **Automatic Optimizations** | | | | | |
| Fusion | ✓ | ✓ | ✓ | ✓ | ✓ |
| Data layout | ✗ | ✗ | ✓ | ✗ | ✓ |
| Tiling | ✗ | ✗ | ✓ | ✓ | ✓ |
| AoS/SoA | ✓ | ✗ | ✓ | ✗ | ✓ |
| **Targets** | | | | | |
| CPU | ✓ | ✓ | ✓ | ✓ | ✓ |
| GPU | ✓ | ✓ | ✓ | ✓ | ✓ |
| FPGA | ✗ | ✗ | ✗ | ✓ | ✗ |
| Clusters | ✗ | ✓ | ✗ | ✓ | ✓ |

The SaC and Futhark implementations are similar in being conventional ahead-of-time compilers that perform a series of compiler passes and optimizations. The passes are largely black-box optimization passes, although SaC provides a variety of command line options to control the process, similar to conventional optimizing compilers. Both Futhark and SaC generate optimized C executables. The Futhark compiler is written in Haskell, while the SaC compiler is written in C. The main difference is that while SaC has an IO facility based on uniqueness types [12], and thus allows the construction of entirely freestanding SaC programs, Futhark does not have any IO facilities, and therefore no way to read input or write output. The output of the Futhark compiler is essentially a *library* that must be invoked by a program written in a general purpose language, through a C-based API. For convenience, the Futhark compiler provides a facility for automatically generating small wrapper C programs for benchmarking and testing purposes.

Accelerate is deeply embedded in Haskell as a Haskell library, and the compiler executes at runtime. Indeed, the Accelerate program is only constructed during the execution of the host Haskell program, where it is then compiled Just-In-Time (JIT) prior to execution. The Accelerate compiler automatically performs many optimizations and rewrites of the program, notably fusion. Although it is, strictly speaking, a JIT compiler, it is largely similar to traditional ahead-of-time compilers, and does not make use of tracing, deoptimization, data-sensitive optimization, or other dynamic behavior common in JIT compilers. It is, however, possible for the user to perform Haskell-level computation to specialize the Accelerate program for a given workload, which can be seen as a form of staged programming. This capability is exploited for some of the benchmarks we use, such as MG (Section 7.4).

DaCe is similar to Accelerate in being embedded in a frontend language (commonly Python), although it differs from the other language implementations in exposing far more control over the optimization process, e.g., with the programmatic or visual tools outlined in Section 4.5.

APL is traditionally interpreted on CPUs, and the Dyalog interpreter provides efficient operations on arrays of different sizes. The Co-dfns APL GPU compiler is written in data parallel APL to facilitate self-hosting on a GPU. It makes heavy use of runtime libraries to achieve performance, often calling specialized routines that implement APL primitives. Where most of the other compilers use sophisticated optimizations that may, however, produce unpredictable performance, the Co-dfns targets predictable performance. It provides tight integration with the Dyalog interpreter, allowing compiled modules to be used "in situ" from within the interpreter, while also producing standalone executables and shared objects that can be linked from other programming languages.

*Hardware targets.* Finally, the language implementations all target multiple architectures, and all target multicores and CUDA/NVIDIA GPUs. The languages may also target other architectures, e.g., APL, DaCe and SaC also target clusters, and DaCe also targets FPGAs.

## 5. Benchmark Rationale, Evaluation Platforms, Methodology

*5.1. Benchmark Rationale*

We illustrate functional array programming using four challenging benchmarks that represent a range of application domains and parallel computational models. Two benchmarks are from public suites, namely MG (numerical aerodynamic simulation) from NAS [41, 42] and Quickhull (computational geometry) from PBBS [43]. The other two are well-known algorithms, namely brute-force $N$-body (astrophysical simulation) and Flash Attention [44] (deep learning).

The benchmarks cover a *variety of code patterns, parallel constructs/structures, and optimization techniques*: Quickhull exercises flattening of irregular nested parallelism, where manual application results in a composition of (all) flat-parallel skeletons—map, scan (prefix sum), reduce-by-index, scatter. The other benchmarks exhibit regular nested parallelism. $N$-body has been described in Section 2. MG is a 27-point stencil solved at different resolutions. Flash Attention has deeply nested structures of parallel and sequential loops that have matrix multiplication solvers at the innermost level.

The benchmarks exercise different optimizations. Quickhull's performance is determined by how well the flat-parallel skeletons are fused and mapped to the hardware. All other benchmarks require careful mapping of computation to different hardware levels, e.g., for achieving temporal reuse from fast memory and for enabling CPU vectorization. Flash Attention requires tiling multiple dimensions into each level of the memory hierarchy. Finally, $N$-body exhibits an interesting case of dataset sensitivity on GPUs: the largest dataset benefits from sequentializing the inner function/loop, while the smaller datasets benefit from parallelizing all functions/loops.

Rather than choosing simple benchmarks that can readily be expressed, and effectively optimized in all five languages, we have selected a set of challenging benchmarks that often take a language outside its "comfort zone". For example, Quickhull utilizes scans (prefix sums), and neither SaC nor DaCe provide a scan primitive. Several benchmarks require nested parallelism, and Accelerate does not normally provide it. MG uses stencils, and Futhark does not provide specific stencil support, nor deep learning support for Flash Attention. The hypothesis is that *collectively* these five research languages offer state-of-the-art solutions to both the expression and performance challenges posed by the benchmarks.

*5.2. Evaluation Platform: the Radboud Server*

The evaluation is conducted on an experimental server at Radboud University that provides both a multicore CPU and a GPU. The server is booked for exclusive use during the experiments. The CPUs are two 16-core AMD EPYC 7313 CPUs running at 3.7 GHz. The peak performance of this machine is 1.895 Tflops (double precision). The RAM has a theoretical peak bandwidth of 204.8 GB/s, though experimentally (STREAM) we observe 140 GB/s.

The server also provides an NVIDIA A30 GPU, with peak performance of 10.3 Tflops and 5.2 Tflops for single and double precision, respectively, and 933 GB/s main memory bandwidth. We use NVCC version 12.1.66 and GCC 11.4.0.

For multicores, we use `numactl --interleave all` to set the page-mapping policy and we configure the thread-binding policy of OpenMP to use `OMP_PLACES="cores" OMP_PROC_BIND=true`.

*5.3. Development Methodology*

The benchmarks are implemented in each of the languages following an iterative procedure: first, we generate a correct implementation, and then repeatedly refactor it to optimize for the target hardware architectures. The final versions of the benchmark code are provided in a public repository[3] and represents the culmination of this process. The optimizing iterations are not described in detail, although we discuss the key optimizations for each language in the Sections 6 to 9.

In contrast to the separate CUDA and OpenMP baselines for each baseline *there is a single benchmark implementation in each language.* This implementation is compiled and optimized for both the multicore and GPU hardware, and for all datasets. The only exception is DaCe, which uses separate CPU and GPU implementations of $N$-body and Flash Attention. We discuss the development effort implications of having a single implementation in Section 10.

*5.4. Measurement Methodology*

We measure total application runtime, but in the case of GPU execution, we exclude the time taken for the (1) CUDA-context initialization, (2) CUDA-kernels compilation, and (3) host-device transfers of the program input and final result (only). For each benchmark and hardware platform, we perform five warm-up runs and then report the average of at least ten runs, or as many as required to reduce standard deviation to less than 3 %.

## 6. Benchmark 1: N-body Simulation

*6.1. Benchmark Description*

The $N$-body simulation benchmark has been described in Section 2.1, together with implementations in the five functional array languages. Recall that each body is associated with multiple values, namely its position and velocity—each represented as a tuple of three double-precision floats—together with its acceleration and mass (both doubles). Hence, there are various ways to represent the bodies in memory, such as using AoS, SoA, or some combination.

---

[3]https://github.com/diku-dk/CFAL-bench/

*6.2. Baselines, Performance Measure, Datasets*

The GPU baseline consists of two code versions, both using AoS layout, whose selection is optimally hardcoded for the target datasets. When the number of bodies $n$ is small ($n \leq 10^4$) we implemented a version that efficiently exploits both levels of parallelism by computing the acceleration for a body with a CUDA block of threads. For larger values of $n$ we use the CUDA-sample implementation[4] that utilizes only the outer level of parallelism. The key optimization is 1D tiling in shared memory. That is, computing the acceleration for a body accesses all other bodies in the same order, so the computation is structured so that a chunk of bodies is collectively copied from global to shared memory by the threads in the same block and then reused. Without tiling, reuse is serviced by the L2 cache, which has higher latency and smaller throughput than GPU shared memory.

We have written a C with OpenMP CPU baseline that (i) parallelizes the computation of the `accel` function across processors using OpenMP and static scheduling, and (ii) ensures that the underlying (GCC) compiler successfully vectorizes the code by exploiting a SoA representation consisting of eight arrays of doubles.

We compute the number of flops for $n$ bodies and $t$ time steps as $(19n^2 + 12n) \cdot t$. Each interaction between two bodies is computed as

$$\frac{m_j(\vec{p_j} - \vec{p_i})}{(\|\vec{p_j} - \vec{p_i}\|^2 + \epsilon^2)^{3/2}} \qquad (6)$$

requiring 16 flops to compute the change in acceleration and another 3 additions to update it. The evaluation uses three datasets corresponding to the following values of $(n,t)$: $(10^3, 10^5)$, $(10^4, 10^3)$ and $(10^5, 10^1)$. Note that these datasets all involve the same amount of work, but vary in how much parallelism is available.

*6.3. N-body Simulation with Futhark*

The Futhark $N$-body simulation is expressed as two nested loops—the outermost a `map` and the innermost a `reduce` (fused with a `map`), as outlined in Section 2.2. Futhark's CPU and GPU backends both generate two code versions: (i) where the parallelism of the `reduce` is exploited (resulting in a segmented reduction), and (ii) where the innermost loop is sequentialised. A code version is dynamically selected at run-time, as explained in Section 4.1. For the $N = 10^6$ case, the outer loop has sufficient parallelism and version (ii) is used.

Futhark's GPU backend tiles the sequentialized inner loop, as in the baseline implementation. This optimization is not implemented for Futhark's CPU backend, which instead depends solely on the hardware caches. The source code uses an AoS layout, but the compiler always transforms it to the SoA layout.

---

[4]github.com/NVIDIA/cuda-samples/tree/master/Samples/5_Domain_Specific/nbody

*6.4. N-body Simulation with Accelerate*

The acceleration between each pair of bodies is calculated by expanding the vector of bodies once along the rows, once along the columns (using `replicate`), and zipping the two resulting matrices with the `accel` function. All rows of the resulting matrix are then summed in parallel using a (rank-polymorphic) `fold`, resulting in a vector. Fusion is essential here as it would be inefficient, both in terms of runtime and memory allocation, to create the intermediate matrices. The fusion pass in the compiler ensures that the resulting code is a single nested parallel loop producing the output vector, without generating any intermediate arrays.

*6.5. N-body Simulation with Single Assignment C*

To support vectorization on the CPU, the SaC *N*-body code uses a struct for the $x, y, z$-coordinates instead of a `double[3]` array. Thus, the compiler eliminates the struct, and the generated code deals with three independent vectors of scalar coordinates. This ensures that the *x*-coordinates of consecutive bodies are contiguous in memory, and likewise for the $y, z$ coordinates and masses.

*6.6. N-body Simulation with APL*

The points and velocities are each stored in an SoA model of nested arrays of 3 elements corresponding to the $x$, $y$, and $z$ coordinates. If the conceptual computation is thought of as a summation of each row of an $N \times N$ matrix, the APL code tiles the matrix into a series of $N \times B$ columns to compute a partial summation of each row a tile at a time. Each tiling pass first computes the differences $d$ for each dimension and then computes the weight matrix $w$ from these differences. To compute the velocity shifts based on these weights and each point's mass $m$, we utilize the expression `(d×w)+.×m[⍵]`, which leverages the `+.×` matrix multiplication in APL.

This particular ordering allows for each tile to first be computed as a fused kernel(s) of the differences and the weight computation, before the fast matrix-multiplication libraries are used in the final computations. This allows the code to make use of the facilities in the APL runtime, since fusion in Co-dfns only occurs on primitives and not on user-defined functions. It also enables portions of the problem to be mapped to fast matrix multiplication code in the runtime.

*6.7. N-body Simulation with DaCe*

For CPU execution, we start with the *N*-body Python program shown in Fig. 7 and optimize the resulting SDFG with the custom workflow summarized in Fig. 9. The data are matrices of size $3 \times N$ and $4 \times N$, corresponding to a SoA representation.

For GPU execution, we use a Python program with two main differences compared with the CPU implementation. The first is additional tiling of one of the inner loops and the use of GPU-shared memory to reduce the number of loads from GPU-global memory. The second is the transposition of the matrices to produce an AoS representation and allow the compiler to vectorize the loads to GPU-shared memory.

*6.8. Summary*

Table 5 summarizes the *N*-body compute rate for the five languages on both multicore CPU and GPU.

Table 5: *N*-body compute rate in Gflops (higher is better) for different numbers of bodies and iterations. The $n = 10^3$ case is done with 16 instead of 32 cores if that gave better results.

|  | $n = 10^3$, $t = 10^5$ | | | $n = 10^4$, $t = 10^3$ | | | $n = 10^5$, $t = 10^1$ | | |
| --- | --- | --- | --- | --- | --- | --- | --- | --- | --- |
|  | CPU | | GPU | CPU | | GPU | CPU | | GPU |
|  | 1C | 32C* |  | 1C | 32C |  | 1C | 32C |  |
| Baseline | 19.0 | 280 | 560 | 18.9 | 583 | 721 | 18.9 | 610 | 1334 |
| Accelerate | 9.3 | 7 | 27 | 9.7 | 261 | 305 | 9.7 | 504 | 303 |
| APL | 1.5 | — | 13 | 1.7 | — | 14 | 1.3 | — | 15 |
| DaCe | 19.0 | 209 | 117 | 19.2 | 560 | 974 | 19.2 | 595 | 1643 |
| Futhark | 18.2 | 127 | 547 | 18.2 | 313 | 968 | 18.1 | 522 | 1576 |
| SaC | 19.4 | 238 | 39 | 19.4 | 556 | 229 | 19.4 | 598 | 264 |

*GPU performance.* Futhark and DaCe beat the GPU baseline on the second and third datasets with speedup factors between 1.18−1.35×. Futhark is competitive with the baseline on the first dataset, but DaCe lags behind because it utilizes only the parallelism of the outer loop, which is insufficient. Accelerate and SaC come next, achieving fractions 5 %, 42 % and 23 % and 7 %, 32 % and 20 % of the baseline performance on the three datasets, respectively; the first dataset similarly suffers due to insufficient utilization of parallelism. APL provides useful acceleration—in that its GPU compute rate is significantly higher than the sequential compute rate—but its performance is only as high as 3.5 % of the GPU baselines.

*Multicore performance.* SaC and DaCe are competitive with the baseline on the second and third datasets (exceeding 95 %), and close on the first dataset (85 % and 75 %). Futhark and Accelerate provide good performance on the third (large) dataset (exceeding 83 %) but perform less well on the second dataset (54 % and 45 %), and even worse on the first (45 % and 2.5 %).

DaCe presents the best blend of performance across the two hardware platforms. SaC is best developed for multicore execution, and its suboptimal GPU performance is directly related to its GPU backend receiving much less attention, and conversely for Futhark and Accelerate. These performance trends are reflected in the other benchmarks.

*Expression.* The baseline consists of three low-level C++ code versions—two aimed at GPU and one at multicore execution—that use *different layouts* (SoA *vs.* AoS) and *manually-applied optimizations* (see Section 6.2). In comparison, the functional array languages use a *single* high-level code (see Section 2) that is *automatically optimized* for multicore and GPU execution. While "beauty is in the eye of the beholder", we consider that all languages provide significantly more elegant code than the baseline—with a shoutout to APL for brevity—and encourage readers to make their own mind by examining our code repository.

## 7. Benchmark 2: MultiGrid (NAS MG)

*7.1. Benchmark Description*

Multigrid methods solve differential equations numerically. The NAS parallel benchmarks [45] provide a simplified version of the V-cycle multigrid method to solve the discrete Poisson problem

$$\nabla^2 u = v \quad (7)$$

on a $n \times n \times n$ grid with periodic boundary conditions. The Laplacian $\nabla^2$ is approximated with a trilinear finite element discretization $A$, which is defined as a function taking and returning an $n \times n \times n$ grid of real numbers:

$$\begin{aligned} A(x))[i_1, i_2, i_3] = \\ \sum_{j_1,j_2,j_3=0}^{2} w[j_1,j_2,j_3] \cdot x[(i_1+j_1-1) \bmod n, \\ (i_2+j_2-1) \bmod n, \\ (i_3+j_3-1) \bmod n] \end{aligned} \quad (8)$$

for some weights $w \in \mathbb{R}^{3 \times 3 \times 3}$. This type of computational pattern is called a *27-point stencil with periodic boundary conditions.*

We define a function `f2c` for projecting a fine grid on a coarser grid. This maps $x$ to $x(1:n:2, 1:n:2, 1:n:2)$, where we use the Fortran notation for the index range from 1 to $n$ with step 2, in each dimension. Similarly, we have a function `c2f`, which prolongates a coarse grid to a fine grid. This embeds some array $x$ of size $n \times n \times n$ into an array $y$ of size $2n \times 2n \times 2n$ as $y(1:n:2, 1:n:2, 1:n:2)$. The remainder of $y$ are zeroes.

The computational core of MG is the V-cycle shown in Algorithm 1. This is a function composed of $A$ and three more 27-point stencils $P, Q, S$. The recursive call causes the stencils to operate on progressively coarser grids, until the input has shape $2 \times 2 \times 2$, after which the grid gets finer again. The MG benchmark, Algorithm 2, repeatedly uses the V-cycle function $M$ to update a solution to Eq. (7).

*7.2. Baselines, Performance Measure, Datasets*

The CPU baseline selected is the Fortran + OpenMP implementation from NAS[5] [41]. The GPU baseline selected is the CUDA implementation of NPB-GPU [6] [42]. Both baselines perform an optimization that exploits a domain-specific property of the $3 \times 3 \times 3$ array of weights, namely that the weights situated at the same distance from the center (at index $[1,1,1]$) are equal, i.e., the 27 weights contain only 4 distinct values. It follows that the number of flops per element can be significantly reduced from the 26 additions and 27 multiplications of a naive implementation, e.g., by performing rewrites of the form $w_1 a + w_1 b + w_1 c + w_1 d = w_1(a+b+c+d)$. Further, the number of additions can also be reduced by observing that for a portion of a slice in the $xy$-plane:

$$\begin{pmatrix} * & - & * \\ - & \circ & - \\ * & - & * \end{pmatrix}$$

the sum of the $*$s and $-$s are used to update not only $\circ$, but also the points directly above and below $\circ$. This can be exploited by precomputing, for each $(i,j)$, two temporary buffers containing the sums $x[i+1,j,.]+x[i-1,j,.]+x[i,j-1,.]+x[i,j+1,.]$, and $x[i+1,j+1,.]+x[i-1,j-1,.]+x[i+1,j-1,.]+x[i-1,j+1,.]$. The temporary buffers are then used to update $x[i,j,.]$. This technique not only reduces the number of operations, but also enables temporal reuse, e.g., the CUDA baseline allocates the temporary buffers in shared memory, thus reducing the number of accesses to global memory.

---

[5]https://www.nas.nasa.gov/software/npb.html
[6]https://github.com/GMAP/NPB-GPU

| **Algorithm 1** The V-cycle function $M$ | **Algorithm 2** MG |
|---|---|
| 1: **procedure** M(r) <br> 2:    **if** $n \neq 2$ **then** <br> 3:       $rs = P(\text{f2c}(r))$ <br> 4:       $zs = M(rs)$ <br> 5:       $z = Q(\text{c2f}(zs))$ <br> 6:       $r = r - A(z)$ <br> 7:       $z = z + S(r)$ <br> 8:    **else** <br> 9:       $z = S(r)$ <br> 10:   **return** z | **Input:** $v$ : an $n \times n \times n$ array. <br> **Input:** $t$ : a number of iterations. <br> **Output:** $u$ : an $n \times n \times n$ array approximately satisfying $\nabla^2 u = v$. <br>    $u = 0$ <br>    **for** $i = 1$ **to** $t$ **do** <br>       $r = v - A(u)$ <br>       $u = u + M(r)$ |

Figure 10: Multigrid method for numerically solving $\nabla^2 u = v$.

We report performance in $Gflops$ following the same methodology as in the NAS benchmark, where the number of flops for an $n \times n \times n$ grid and $t$ iterations is considered to be $58tn^3$. The evaluation uses the $ClassA$, $ClassB$, and $ClassC$ standard NAS datasets.

*7.3. MultiGrid with Futhark*

The Futhark implementation supports both the naive and optimized versions of MG introduced in Sections 7.1 and 7.2. A simplified code excerpt from the latter is shown in Fig. 11. In contrast to the lengthy baseline implementation (4662 source lines of code in Table 4) that consists of a multitude of distinct (specialized) computational kernels, the Futhark implementation abstracts (most of) the core computation of the 27-point stencil into *one* function that is

```
1   def relaxNAS (n:i64)                              def getEachElm arr i j k =
2               (elmAt:i32->i32->i32->real)             #[unsafe] arr[i, j, k]
3               (ws: [4]real) : *[n][n][n]real =
4    let computeRow (ii: i64)(jj: i64) : [n]real =   def get8thElm arr i j k =
5      let (i, j) = (i32.i64 ii, i32.i64 jj)           if (i%2) + (j%2) + (k%2) == 3
6      let f (k: i32) =                                then #[unsafe]
7        ( elmAt i ((j-1) % n) k +                         arr[i/2,j/2,k/2]
8          elmAt i ((j+1) % n) k +                     else 0.0
9          elmAt ((i-1) % n) j k +
10         elmAt ((i+1) % n) j k                     def Q [n] relaxKer
11       ,                                                   (zs: [n][n][n]real)
12         elmAt ((i-1) % n) ((j-1) % n) k +               : [2*n][2*n][2*n]real =
13         elmAt ((i-1) % n) ((j+1) % n) k +         relaxKer (2*n) (get8thElm zs)
14         elmAt ((i+1) % n) ((j-1) % n) k +                 [1, 1/2, 1/4, 1/8]
15         elmAt ((i+1) % n) ((j+1) % n) k
16       )                                           def A [n] relaxKer
17     let tab n h = map (h<-<i32.i64) [0...n-1]              (z: [n][n][n]real)
18     let (u1s, u2s) = tab n f                               (r: [n][n][n]real)
19     let g u1s u2s (k: i32) = #[unsafe]                   : [n][n][n]real =
20             ws[0] * elmAt i j k                   relaxKer n (getEachElm z)
21           + ws[1] * (u1s[k] +                             [-8/3, 0, 1/6, 1/12]
22                      elmAt i j ((k-1) % n) +      |> map2 (map2 (map2 (-))) r
23                      elmAt i j ((k+1) % n))
24           + ws[2] * (u2s[k] + u1s[(k-1) % n] +    def S [n] relaxKer
25                               u1s[(k+1) % n])             (ws: [4]real)
26           + ws[3] * (u2s[(k-1)%n] + u2s[(k+1)%n])         (r: [n][n][n]real)
27     in  tab n g                                            (z: [n][n][n]real)
28   in  map (\i -> map (computeRow i) (0...n-1))           : [n][n][n]real =
29           (0...n-1)                               relaxKer n (getEachElm r) ws
30                                                   |> map2 (map2 (map2 (+))) z
```

Figure 11: The left-hand side shows a *simplified* implementation of the optimized computational kernel of MG, named relaxNAS, which is made generic by parameterizing it over an index function (elmAt) rather than a 3D array. The right-hand side shows the instantiations required to implement the computations $Q(\mathtt{c2f}(zs))$, $r - A(z)$ and $z + S(r)$ in Algorithm 1. Of note, the actual implementation uses fast arithmetic for modulo operations. #[unsafe] prevents the insertion of dynamic bounds checks (which are expensive on GPUs).

suitably parameterized, e.g., over (i) the size of the result-array dimension $n$, (ii) an index-function representation *elmAt* of the source array and (iii) the array of weights *ws*. The instantiations required to compute $Q(\mathtt{c2f}(zs))$, $r - A(z)$ and $z + S(r)$ in Algorithm 1 are shown on the right-hand side of Fig. 11, as functions named simply $Q$, $A$ and $S$, respectively. Finally, the function parameter *relaxKer* may be instantiated with either the naive or optimized version (*relaxNAS*). On the cons side, Futhark does not support arbitrary recursion leading to a tedious (and ugly) implementation of procedure $M$ (of Algorithm 1) where recursion is modeled by a while loop that explicitly maintains the stack by means of updates to a jagged array.

The high-level implementation critically relies on the compiler to specialize the code. For example, Futhark's simplification engine aggressively inlines functions such as *relaxNAS*, *getEachElm*, *get8thElm*, $f$, $g$ at call sites, and performs a battery of transformations, such as copy propagation, constant folding, common-subexpression and dead-variable elimination, and scalar specialization of array-literal indexing. For example, the specialization of *relaxNas* at the call site of $A$ (in Fig. 11), will eliminate the whole term starting with $ws[1] * (\ldots)$ appearing in the computation of $g$ (at lines 21-23)—because $ws[1]$ corresponds

```
double[d:shp] stencil(double[d:shp] x, double[d:wshp] w)
{
  return {i -> sum({j -> w[j] * x[mod(i + j - wshp / 2, shp)]})
            | i < shp};
}
```

Figure 12: A Rank-Polymorphic stencil in SaC. For suitably chosen weights w, this computes the scaled (higher order) derivative of a discretized function x.

to constant 0 in this context, and multiplication with zero always results in zero.

Furthermore, as presented in Section 4.1, for the GPU case, the incremental flattening transformation will generate an "intrablock" code version that essentially maps the (inner) parallelism available within *computeRow* to the CUDA block level and the outer parallelism to the CUDA grid level, that is, the two nested *map*s at line 28. Consequently, the arrays $u1s, u2s$ produced by *tab n f* (line 18) are placed in shared memory and reused from there in the computation of *tab n g* (line 27). Finally, memory-related optimizations eliminate some of the array copying operations resulting from explicit modeling of the (recursion) stack.

### 7.4. MultiGrid with Accelerate

The Accelerate implementation uses recursion to implement procedure M, despite Accelerate, like Futhark, not supporting recursion. It does so by exploiting the recursion provided by the Haskell host language via meta-programming. This unfolds the recursion, similar to how loop unrolling converts a loop to a sequence of instructions.

Accelerate features stencil primitives that simplify the implementation of relaxations. Again, using meta-programming, the weights are inlined into the stencil kernel.

Accelerate's fusion algorithm fuses `f2c` and `c2f` into a following stencil operation. This improves performance, while allowing the user to write the algorithm at a higher level.

### 7.5. MultiGrid with Single Assignment C

SaC implements MultiGrid recursively, faithful to the specification. We use the generic stencil implementation shown in Fig. 12. This function generalizes to arbitrary dimension d and weights of arbitrary shape wshp. The modulo computations (mod) are optimized away by generating multiple partitions.

SaC's constant folder currently reduces the number of multiplications automatically, but not the number of additions. The Fortran version has one temporary array per thread, which it reuses. As SaC does not have explicit parallelism, getting this to happen is not straightforward. An alternative would be to compute the temporary arrays for all $(i, j)$ in parallel, but that is not worth it for cache reasons. The GPU code generation suffers greatly from a problem in the analysis that prevents arrays from being moved from and to the device.

### 7.6. MultiGrid with APL

The APL implementation is a fairly direct translation of the pseudo-code:

```
M←{α≤1:S st ω ◊ z+S st ω-A st⊢z←Q st up(α-1)∇dn P st ω}
L2←{((+/,ω×ω)÷×/ρω)*.5}
MG←{⎕L2 α-A st α{⎕ω+k M α-A st ω}⍣IT⊢ω}
```

Where *MG* is run for *IT* iterations and the functions *st*, *dn*, and *up* correspond to the stencil computation, downsampling, and upsampling, respectively. Downsampling and upsampling are implemented as direct array indexing operations:

```
dn←{ω[i;i;i←1+2×ι(≢ω)÷2]}              ⍝ Downsample
up←{X⊣X[i;i;i←1+2×ι≢ω]←ω⊣X←1e¯100⍴⍨2×ρω}  ⍝ Upsample
```

APL does have a built-in stencil operator, but it is not suitable for high-performance optimizations such as those available in MG. We instead implement a custom stencil function *st* that minimizes the number of arithmetic operations as described in Section 7.2. Since the entire function can be expressed as a single composition of arithmetic primitives and rotations, the resulting code can be fused into a single conceptual kernel:

```
⍝ MG Stencil Operation
st←{k←α
    x←ω×k[0]
    x+←k[1]×(1⊖ω)+(¯1⊖ω)+r0←(1⌽ω)+(¯1⌽ω)+r01←(1⌽[1]ω)+¯1⌽[1]ω
    x+←k[2]×(1⊖r0)+(¯1⊖r0)+r1←(1⌽r01)+¯1⌽r01
    x+←k[3]×(1⊖r1)+¯1⊖r1
    ⎕x
}
```

### 7.7. MultiGrid with DaCe

Since the SDFG IR does not support recursion, DaCe's MG implementation is based on a Python version [46]. We automatically optimize the CPU and GPU implementations, which fuses parallel computations (Map scopes) and tiles those that contain write-conflict resolutions to minimize atomic operations. However, as the auto-optimizer transforms the graph in a specific order, some optimization opportunities may be missed. For example, Fig. 13 shows the MG's `norm2u3` method and its DaCe-Python implementation, which includes two Map-Reduce patterns for computing `s` and `rnmu`. The auto-optimizer successfully fuses these computations into a single Map scope but cannot subsequently tile it because the relevant DaCe optimization does not support Map scopes with write (WCR) conflicts to multiple data containers. To work around this limitation, we design a custom workflow. First, we apply the `MapReduceFusion` transformation repeatedly until no subgraph matches the optimization pattern, resulting in two fused Map scopes, one for `s` and one for `rnmu`. Only then do we apply WCR tiling, which matches since each scope writes to a single data container.

```
1   n0i, n0j, n0k = \
2     (dace.symbol(s, dtype=dace.int32) for s in ('n0i', 'n0j', 'n0k'))
3
4   @dace.program
5   def norm2u3_dace(r: dace.float64[n0i, n0j, n0k], dn: dace.int32):
6       s = dace.reduce(lambda a, b: a + b,
7                       r[1:-1, 1:-1, 1:-1] * r[1:-1, 1:-1, 1:-1],
8                       identity=0.0)
9       rnmu = dace.reduce(lambda a, b: max(a, b),
10                          numpy.abs(r[1:-1, 1:-1, 1:-1]),
11                          identity=0.0)
12      rnm2 = numpy.sqrt(s / dn)
13      return rnm2, rnmu
14
15  fast_sdfg = norm2u3_dace.to_sdfg()
16  auto_optimize(fast_sdfg)
17
18  faster_sdfg = norm2u3_dace.to_sdfg()
19  faster_sdfg.apply_transformations_repeated([MapReduceFusion])
20  tile_wcrs(faster_sdfg)
21  greedy_fuse(faster_sdfg)
```

Figure 13: DaCe Python implementation of norm2u3 and its optimization.

### 7.8. Summary

The NAS MG performance results for the five languages on both multicore CPU and GPU are summarized in Table 6. We follow the baseline performance measurement and report compute rate (Gflops).

Table 6: NAS MG compute rate in Gflops (higher is better) for NAS classes A, B, C.

|  | Class A | | | Class B | | | Class C | | |
|---|---|---|---|---|---|---|---|---|---|
|  | CPU | | GPU | CPU | | GPU | CPU | | GPU |
|  | 1C | 32C |  | 1C | 32C |  | 1C | 32C |  |
| Baseline | 8.7 | 83 | 185 | 9.0 | 95 | 240 | 8.2 | 48 | 240 |
| Accelerate | 1.6 | 7 | 72 | 1.7 | 8 | 71 | — | 14 | 82 |
| APL | 0.5 | — | 7 | 0.5 | — | 7 | 0.5 | — | 7 |
| DaCe | 3.3 | 28 | 129 | 3.6 | 41 | 141 | 3.0 | 37 | 227 |
| Futhark | 1.2 | 16 | 235 | 1.4 | 22 | 237 | 1.4 | 21 | 238 |
| SaC | 5.9 | 43 | — | 6.6 | 53 | — | 5.0 | 39 | — |

*GPU performance.* Futhark offers constant performance across datasets ($235 - 238$ Gflops), which is competitive with the baseline for the two larger B/C datasets ($99\,\%$) and outperforms the baseline on the smallest A dataset ($127\,\%$). DaCe also offers competitive performance on the largest C dataset ($95\,\%$), but is less competitive for the first two datasets ($70\,\%$ and $59\,\%$). Accelerate offers relatively constant performance across datasets ($71 - 82$ Gflops) reaching $30\,\%$ to $39\,\%$ of the baseline performance. APL reports useful acceleration in comparison with the sequential execution, but reaches only $3\,\%$ to $4\,\%$ of the baseline performance. For SaC, compiler shortcomings have prevented GPU execution.

| **Algorithm 3** QuickHull | **Algorithm 4** Divide-And-Conquer Helper |
|---|---|
| **Input:** $S$: a set of $n \geq 2$ two-dimensional points | 1: **procedure** FINDHULL($S$, $P$, $Q$) |
| **Output:** $CH$: the convex-hull set of $S$ | 2:     **if** $S \neq \emptyset$ **then** |
|     $(A, B)$ = the leftmost and rightmost points of $S$ | 3:         $C$ = furthest point of $S$ from line $PQ$ |
|     $S_{1,2}$ = points of $S$ above and below line $AB$ | 4:         $(S_l, S_r)$ = the points of $S$ on the left- |
|     $CH = \{A, B\} \cup$ | 5:            and right-hand side of lines |
|       FINDHULL($S_1$, A, B) | 6:            $CP$ and $CQ$, respectively |
|       $\cup$ | 7:            (and not inside $\Delta PCQ$) |
|       FINDHULL($S_2$, A, B) | 8:         $Hull = \{C\} \cup$ |
| | 9:            FINDHULL($S_l$, P, C) $\cup$ |
| | 10:           FINDHULL($S_r$, C, Q) |
| | 11:     **else** |
| | 12:         $Hull = \emptyset$ |
| | 13:     **return** $Hull$ |

*Multicore performance.* The baseline outperforms all of the functional array languages, likely due to suboptimal thread-binding policies. SaC performs best, reaching 52 %, 56 % and 81 % of baseline performance on the `A,B,C` datasets, followed by DaCe with 34 %, 43 % and 77 %. At a considerable distance come Futhark (19 %, 23 % and 44 %) and Accelerate (8 %, 8 % and 29 %), whose multicore backends are less mature than the GPU backends. Notably, Futhark exhibits very poor sequential performance but excellent scalability, albeit the latter is likely a case of "scaling the overhead".

*Expression.* The Accelerate, APL, Futhark, and SaC teams found it infeasible to develop their code by following the long, low-level, and complex baseline code (4662 source lines of code, Table 4). The baseline specializes and hand-optimizes one generic stencil computation for a multitude of contexts. Instead, our implementation teams based their implementations on either the NAS specification or the SaC code, which is short, clean, and arguably very close to the algorithmic formulation (Sections 7.1 and 7.2). Moreover, Fig. 11 shows that the key baseline optimizations, providing temporal reuse and a reduced number of arithmetic operations, can also be expressed generically and concisely without compromising GPU performance, as evidenced by Futhark's performance. While code comparisons are subjective, we invite the reader to reach a view by browsing our code repository [7].

# 8. Benchmark 3: Quickhull

## 8.1. Benchmark Description

The Quickhull benchmark computes the convex hull of a finite set of points in 2-dimensional space. Algorithm 3 shows the pseudocode of the Quickhull algorithm: given a set $S$ containing at least 2 two-dimensional points, the convex hull of $S$ is computed by first (i) identifying the leftmost and rightmost points of $S$, named $A$ and $B$, respectively, then (ii) by partitioning $S$ into subsets $S_1$ and $S_2$ corresponding to the points below and above the line $AB$, and finally (iii) by combining the convex hulls of $S_1$ and $S_2$. The last step is facilitated by the $findHull$ helper function presented in Algorithm 4. The divide-and-conquer $findHull$ function takes two points $P$ and $Q$ and a set of points $S$ that are either all above or all below line $PQ$. $findHull$ computes the point $C$ of $S$ that is furthest away from $PQ$, and hence must belong to the convex hull, then it eliminates from $S$ the points inside triangle $\Delta CPQ$ as they are guaranteed not to belong to the convex hull. The resulting set is partitioned into the points that are on the right- and left-hand side of lines $CP$ and $CQ$, named $S_l$ and $S_r$, respectively. The convex hull of $S$ consists of point $C$ together with the convex hulls of $S_l$ and $S_r$.

Such divide-and-conquer recursion induces *irregular nested parallelism*: the recursive calls form a binary tree where each call performs parallel operations (e.g., filtering, partitioning) on arrays of different sizes. Importantly, all the nodes at the same level in the tree can be processed in parallel.

Ideally, the Quickhull implementation should resemble the divide-and-conquer structure of Algorithm 4. This approach was pioneered in the NESL language [10], and can be automatically mapped to hardware using flattening transformations [20], and may target multicore [47] or GPU architectures [48, 49]. Since none of the studied languages support flattening of irregular parallelism, their corresponding implementations are derived by manually flattening the code corresponding to Algorithm 4.

## 8.2. Baselines, Performance Measure, Datasets

The CPU baseline selected is the convex-hull implementation of the PBBS benchmark[8] [43], that we believe to offer state-of-the-art multicore performance. The implementation uses a fork-join (task-based) strategy that dynamically decomposes and exploits both levels of parallelism while tasks have good granularity. This strategy results in cache-friendly behavior that cannot be replicated by any of the studied languages, given their data-parallel paradigm. This paradigm effectively enforces barriers around the parallel processing of the nodes at each level in the tree, resulting in larger re-use distances.

We do not use a GPU baseline as, other than the studied languages, we could not find a publicly available implementation competitive with the PBBS implementation. This emphasizes the importance of high-level data-parallel models offering efficient hardware mappings, e.g., to GPUs.

---

[7]https://github.com/diku-dk/CFAL-bench/
[8]https://github.com/cmuparlay/pbbsbench

We follow PBBS in reporting performance as runtime. The evaluation uses three PBBS datasets, each consisting of a set of one hundred million points that are uniformly distributed in the interior of a *Rectangle*, in the interior of a *Circle* (a disk), and on a *Quadratic* curve. Their convex hulls consist of 49, 1681, and 548136 points, respectively.[9]

*8.3. Quickhull with Futhark*

The Futhark implementation of Quickhull has been manually flattened to be expressible with flat data parallelism. At a high level, it consists of an outer `while` loop that tracks the points determined to be part of the hull, as well as a sequence of sets of points that have not yet been classified. This sequence is conceptually a multidimensional jagged array, and is represented using a standard flag array approach. The implementation consists of a composition of map, reduce-by-index, filter, and partition operations, where the implementation of the latter two relies on scan (a.k.a., prefix sum) and scatter (a.k.a., parallel write). Efficient execution on GPU hardware is enabled by specialized code generation for scan and reduce-by-index (Section 4.1).

*8.4. Quickhull with Accelerate*

The Accelerate implementation, like Futharks', is also flattened manually, making it necessary to maintain some segment information. The main work during one step is a parallel filter operation, applied to all segments. This consists of a mapping operation to determine which points to discard, a segmented scan over the resulting flag vector to compute the target indices, and a subsequent permutation.

*8.5. Quickhull with APL*

The APL version first reorders the points so that membership in the hull set can be tracked using a Boolean mask to partition the data vectors into segments. On each iteration, this mask is used to group each point and determine membership within the hull set. Points falling within the hull are removed, while the new point that divides each segment is computed and added to the set by flipping its bit in the bitmask.

APL currently lacks a method for computing Scans or Reductions over subsegments of an array indicated by a key. Instead, the expression `M[R≠⍋K[M←⍋D]]` computes the first indices of *D* that have the maximum *D* for each unique key in *K*. This expression is reasonably efficient, but involves sorting two vectors to find the max, which is significantly less efficient than a max reduction might be. Moreover, the above code does not split out the hull elements from elements still to be determined; as a result, if the number of points on the hull is large, we may redundantly compute distances that do not matter over time, leading to reduced performance when the hull contains many points. Finally, reordering the points at the beginning instead of at the end requires sorting a much larger array.

---
[9]The datasets are named as in PBBS and are generated with the PBBS infrastructure.

*8.6. Quickhull with SaC*

We implement the algorithm recursively. As SaC's syntax is very close to C, we can straightforwardly adept the partition scheme from Edelkamp [50]. We also find the furthest points in the same loop that partitions $S$ into $S_l$ and $S_r$ to reduce the number of passes over the data. As PBBS performs a Hoare partition and does not merge these loops, the SaC implementation is at least twice as fast sequentially. PBBS also works with indices instead of permuting the array, which gives bad cache behaviour for deep recursions. For this reason SaC is even five times faster for the Quadratic data set. The compiler is not able to parallelize recursive calls. We are not able to implement the flattened algorithm efficiently as SaC lacks parallel writes. This means we have no parallel results.

*8.7. QuickHull with DaCe*

DaCe does not support recursion and, therefore, cannot directly implement Algorithm 4. The alternative formulation using flattening transformations, described in Section 8.1, could be implemented in DaCe by composing control flow (for-loops) with map scopes and reductions to compensate for the lack of (segmented) scan primitives. However, such an implementation would be tedious and uncompetitive with languages that can express those parallel paradigms naturally. The Library Node mechanic could also be leveraged to map a high-level (segmented) scan representation directly to optimized calls, e.g., to the NVIDIA Thrust library for GPUs. Overall we consider the additions required to facilitate a reasonable QuickHull implementation in DaCe to be beyond the scope of this work.

*8.8. Summary*

It is not feasible to develop parallel Quickhull implementations in DaCe or SaC, mainly because they lack support for a key operation like prefix-sum or scatter, or for fork-join parallelism. The Quickhull performance results of Accelerate, APL, and Futhark on multicore CPU and GPU are summarised in Table 7. We follow the baseline performance measurement and report runtime (in seconds).

*GPU performance.* Futhark outperforms the multicore baseline, achieving 313 %, 234 % and 430 % of its performance. Accelerate is competitive with the baseline on the Circle (125 %) and Rectangle (96 %) datasets, but falls behind on Quadratic (68 %). While APL reduces runtime compared with its sequential performance, it reaches only 16 %, 16 % and 39 % of the baseline performance on Circle, Rectangle, and Quadratic datasets.

Table 7: Quickhull runtime in seconds (lower is better) for $10^8$ points sampled uniformly from three geometric shapes.

|  | Circle | | | Rectangle | | | Quadratic | | |
|---:|:---:|:---:|:---:|:---:|:---:|:---:|:---:|:---:|:---:|
|  | CPU | | GPU | CPU | | GPU | CPU | | GPU |
|  | 1C | 32C |  | 1C | 32C |  | 1C | 32C |  |
| Baseline | 4.4 | 0.2 | — | 3.4 | 0.11 | — | 35.1 | 2.9 | — |
| Accelerate | 7.4 | 1.6 | 0.160 | 3.6 | 1.18 | 0.114 | 48.4 | 12.5 | 4.28 |
| APL | 22.2 | — | 1.220 | 14.9 | — | 0.690 | 113.0 | — | 7.57 |
| DaCe | — | — | — | — | — | — | — | — | — |
| Futhark | 5.6 | 1.3 | 0.064 | 3.8 | 1.15 | 0.047 | 37.6 | 4.0 | 0.68 |
| SaC | 1.8 | — | — | 1.6 | — | — | 7.0 | — | — |

*Multicore performance.* The Accelerate and Futhark implementations are obtained by manually applying a flattening transformation [20] to the divide-and-conquer (irregular) parallelism exposed by Quickhull. The baseline outperforms all languages as it uses a fork-join style that dynamically spawns parallel tasks that have good granularity and temporal reuse (Section 8.2). Futhark achieves 16 %, 10 % and 72 % and Accelerate 13 %, 9 % and 23 % of the baseline performance on the three datasets, respectively.

*Expression.* It seems that Quickhull is best expressed in a divide-and-conquer parallel style, similar to Algorithm 4, where the text at lines 3 and 5–8 is implemented with data-parallel constructs such as map-reduce and partition. This would offer a clean and algorithm-faithful high-level specification that preserves the opportunities for efficient mapping to different architectures—e.g., by specializing the flattening transformation for multicore [47] or GPU [48, 49] execution. In principle, this would produce (1) efficient multi-core code resembling the baseline implementation and (2) efficient GPU code resembling the one of Accelerate/Futhark. We consider that our implementations are shorter, easier to understand and more high-level than the baseline. For example, they contain no notion of memory or memory management, and provide correct-by-construction parallelism enabled by implicitly-parallel constructs.

## 9. Benchmark 4: Flash Attention

### 9.1. Benchmark Description

Self-Attention is a machine-learning mechanism for finding dependencies among inputs in a sequence [51]. We consider a sequence of $N$ words or tokens, each with an embedding $\mathbf{v}_i, i \in 1..N$ of length $M$. We want to find another embedding $\mathbf{o}_i$ for each word that encodes its relationship to the other tokens. To that end, we select one word, the query, and we compute its similarities $\mathbf{s}_{ij}, j \in 1..N$ to each other token as the dot product of their embeddings, $\mathbf{v}_i \cdot \mathbf{v}_j$. We normalize these similarities using Softmax [52], converting them to a probability distribution. Multiplying the result vector $\mathbf{p}_i$ with the initial word embeddings produces the final result.

| **Algorithm 5** Standard Self-Attention | **Algorithm 6** Custom Attention |
| --- | --- |
| **Input: Q** : $N \times d$ matrix of query weights. | **Input:** same as Algorithm 5 |
| **Input: K** : $N \times d$ matrix of key weights. | **Output:** same as Algorithm 5 |
| **Input: V** : $N \times d$ matrix of value weights. | 1: **for all** $i \in 0 \mathrel{..} N-1$ **by** $d$ **do** |
| **Output: O** : $N \times d$ attention matrix. | 2: $\quad$ **Qb** $= \mathbf{Q}_{i:i+d,\ :}$ |
| 1: $\mathbf{S} = \mathbf{Q} \times \mathbf{K}^T$ | 3: $\quad$ **Sb** $= \mathbf{Qb} \times \mathbf{K}^T$ |
| 2: $\mathbf{P} = softmax\,(\mathbf{S})$ | 4: $\quad$ **Pb** $= softmaxOnline(\mathbf{Sb})$ |
| 3: $\mathbf{O} = \mathbf{P} \times \mathbf{V}$ | 5: $\quad \mathbf{O}_{i:i+d,\ :}\ =\ \mathbf{Pb} \times \mathbf{V}$ |

Figure 14: In the code, $\mathbf{K}^T$ denotes the transpose of $\mathbf{K}$ and $\times$ denotes matrix multiplication. **forall** $i \in 0 \mathrel{..} N-1$ **by** $d$ denotes a parallel (map-like) computation in which $i$ starts from 0 and advances with a step of $d$, i.e., $i = 0, d, 2 \cdot d, \ldots$. **Qb** iterates over the $d \times d$ slices of **Q** and the last line of Algorithm 6 computes the corresponding slice of the result **O**.

Machine learning architectures, like transformers, learn weights for the different embedding uses: **Q** for query words, **K** for the similarity computation, and **V** for the output embedding. In Multi-head Attention, multiple heads compute the attention corresponding to distinct parts of the word embeddings in parallel, each with size $d$. Hence, in Algorithm 5, the weight matrices have size $N \times d$. Standard Self-Attention materializes all of the **S** and **P** matrices, which have size $N \times N$. In applications of interest, the sequence is large enough to consider $N \gg d$, and $Q$, $K$, $V$, and $O$ are all tall and skinny matrices. Multiplying such matrices exhibits low arithmetic intensity (computation is $O\left(N^2 d\right)$, but memory operations are $O\left(N^2\right)$), making self-attention memory-bound.

Softmax, shown in Algorithm 7, normalises the similarities matrix **S** by computing for every row $i$ two quantities: the maximum value $m_i$ and the normalisation term $l_i$, defined as $l_i = \sum_{j=1}^{N} \exp(s_{i,j} - m_i)$. The output probabilities **P** are computed as $p_{i,j} = \exp(s_{i,j} - m_i)/l_i$ to avoid overflow. Overall, each input row must be read three times since the dependency on $m_i$ disallows loop fusion. An alternative formulation called Online Softmax [53] and shown in Algorithm 8 removes this constraint by using exponent rules to compute $l_i$ recursively, reducing the required reads of the input and enabling computation of the output in blocks.

Flash Attention [44] utilizes Online Softmax to tile the matrix multiplications and the normalization operation, fuse them together, and perform the overall computation in $T_i \times T_j$ blocks. In this reformulation, the intermediate results are small enough to fit in fast memory (e.g., registers and CUDA shared memory), dramatically reducing accesses to slower global memory. An implementation of the algorithm in DaCe data-centric Python is shown in Fig. 17.

Finally, Algorithm 6 presents a midpoint between Standard and Flash Attention, dubbed Custom Attention, that blocks the computation over $d \times d$ slices of $Q$ and computes corresponding slices of the result **O**. Custom Attention does not serialize computation, but also does not guarantee that the intermediate matrices (**Sb** and **Pb** of size $d \times N$) can be maintained in fast memory.

| **Algorithm 7** Stable Softmax | **Algorithm 8** Online Softmax |
|---|---|
| **Input: S**: $M \times M$ input matrix. | **Output: P**: $M \times M$ output matrix. |
| 1: **for all** $i \in 0 \mathrel{{.}\,{.}} M-1$ **do** | 1: **for all** $i \in 0 \mathrel{{.}\,{.}} M-1$ **do** |
| 2: | 2: $\quad m_i = -\infty, l_i = 0$ |
| 3: | 3: $\quad$ **for** $jj \in 0 \mathrel{{.}\,{.}} M-1$ **by** $T$ **do** |
| 4: | 4: $\quad\quad \tilde{s} : \text{vect}[T] = copy(S_{i,jj:jj+T})$ |
| 5: | 5: $\quad\quad m_i'' = \overline{max}(\tilde{s})$ |
| 6: $\quad m_i = \overline{max}(S_{i,:})$ | 6: $\quad\quad m_i' = \max(m_i,\ m_i'')$ |
| 7: $\quad \tilde{p} : \text{vect}[M]$ | 7: $\quad\quad \tilde{p} : \text{vect}[T]$ |
| 8: $\quad$ **for all** $j \in 0 \mathrel{{.}\,{.}} M-1$ **do** | 8: $\quad\quad$ **for all** $j \in 0 \mathrel{{.}\,{.}} T-1$ **do** |
| 9: $\quad\quad \tilde{p}_j = \exp(S_{i,j} - m_i)$ | 9: $\quad\quad\quad \tilde{p}_j = \exp(\tilde{s}_j - m_i')$ |
| 10: | 10: $\quad\quad l_i' = \exp(m_i'' - m_i') \cdot \overline{sum}(\tilde{p})$ |
| 11: $\quad l_i = \overline{sum}(\tilde{p})$ | 11: $\quad\quad l_i = l_i' + l_i \cdot \exp(m_i - m_i')$ |
| 12: $\quad$ **for all** $j \in 0 \mathrel{{.}\,{.}} M-1$ **do** | 12: $\quad\quad m_i = m_i'$ |
| 13: $\quad\quad P_{i,j} = \exp(S_{i,j} - m_i)/l_i$ | 13: $\quad$ **for all** $j \in 0 \mathrel{{.}\,{.}} M-1$ **do** |
| | 14: $\quad\quad P_{i,j} = \exp(S_{i,j} - m_i)/l_i$ |

Figure 15: Algorithm 8 uses the same input and output as Algorithm 7. $\overline{max}$ and $\overline{sum}$ denote reductions of their vector argument. $\tilde{p}$ : vect[$X$] declares of a temporary vector $\tilde{p}$ of length $X$. **for all** denotes parallel (map-like) computations and **for** $jj \in 0 \mathrel{{.}\,{.}} M-1$ **by** $T$ denotes a sequential computation in which the loop index $jj$ advances with step $T$. Online Softmax is an algorithmic refinement that reorganizes the computation as a sequence of parallel computations of length $T$, where $T$ is a configurable tile size. This significantly reduces the number of accesses to global memory because vectors $\tilde{s}$ and $\tilde{p}$ now fit in scratchpad memory.

## 9.2. Baselines, Performance Measure, Datasets

The reference Flash Attention implementation [54] supports half-precision arithmetic (fp16 and bf16) and utilizes the specialized tensor (matrix) core units on NVIDIA (AMD) GPUs. Since not all languages support half-precision arithmetic or execution on tensor cores, we write our own GPU baseline using single-precision arithmetic and the regular FPUs. Following the high-level algorithm in Fig. 17, each $T_i \times d$ block of the output **O** is assigned to a single thread block. The computation is done in a single kernel in $T_j$ steps and utilizes GPU-shared memory and registers to reduce loads from GPU-global memory. Each thread block reads its assigned **O** and **Q** blocks and the whole **K** and **V** matrices. Therefore, **O** and **Q** are read only once, while **K** and **V** are loaded as many times as the number of thread blocks. All intermediate results, for example, the maximum values $m$ and normalization factors $l$, are written to shared memory and registers. The implementation further 2-D tiles the computation of each thread block and utilizes memory coalescing to achieve high GPU-global memory throughput.

The CPU baseline tiles the computation in the same way. Each $T_i \times d$ block of the output **O** is assigned to a single CPU thread. The matrix multiplications are executed with a specialized single-threaded implementation optimized for the AMD Zen architecture.

```
1  def mmmT [m][n][k] (as: [m][k]f32)(bs: [n][k]f32) : [m][n]f32 =
2    map (\a -> map (\b -> map2 (*) a b |> f32.sum) bs) as
3  def mmm [m][n][k] (as: [m][k]f32) (bs: [k][n]f32) : [m][n]f32 =
4    mmmT as (transpose bs)
5
6  def FlashAttention [d][m] (Q: [m][d][d]f32) (K: [m*d][d]f32)
7                            (V: [m*d][d]f32) : [m][d][d]f32 =
8    let mkOneBlock Qi =
9          let P = mmmT Qi K
10              |> onlineSoftmax
11         in  mmm P V
12   in  map mkOneBlock Q
```

Figure 16: Custom Attention in Futhark. The implementation of onlineSoftmax (not shown) corresponds to Algorithm 8

We report the performance in *Gflops* and consider only the operations present in the standard Self-Attention algorithm when measuring the workload, i.e., $N^2(4d+5)$ floating point operations. Our evaluation uses four randomly generated datasets with the following parameters: (1) $d = 64, N = 16,384$, (2) $d = 64, N = 32,768$, (3) $d = 128, N = 8,192$, and (4) $d = 128, N = 16,384$.

*9.3. Flash Attention with Futhark*

Futhark's implementation in Fig. 16, closely resembles Custom Attention, Algorithm 6. During GPU compilation the incremental flattening transformation distributes the outer map processing each $d \times d$ block of $Q$ (line 12) across the computations of (i) the transposed matrix multiplication mmmT at line 9, (ii) the Online Softmax of Algorithm 8 at line 10, and (iii) the second matrix multiplication mmm at line 11. The resulting batch matrix multiplication kernels (i & iii) are optimized for spatial and temporal locality using block and register tiling. Online Softmax (ii) is similarly decomposed into an intragroup kernel that performs the computation of $m_i$ and $l_i$ in shared memory, and another kernel that updates the elements of $P$.

The CPU compilation diverges early from the GPU compilation and does not perform locality optimizations, resulting in very poor performance. Notably, the compilation of Custom Attention manifests the intermediate $P$ matrix in global memory, and hence cannot match the performance of the baseline that implements the more efficient Flash Attention algorithm.

*9.4. Flash Attention with Accelerate*

Accelerate implements Custom Attention, Algorithm 6. We find that the performance is sensitive to the choice of block size. Matrix multiplication is implemented naively in the benchmarks. A faster alternative would be to use the matrix multiplication available through Accelerate's BLAS binding.

*9.5. Flash Attention with Single Assignment C*

SaC implements Algorithm 6, but does not yet provide register tiling. The GPU backend is not able to generate a kernel for all computations. Moreover, the code does not vectorize on the CPU. That would require a non-scalar fold, which we cannot implement efficiently.

```
1  @dace.program
2  def flash_attention_dace(Q: dace.float32[N, d], K: dace.float32[N, d],
3                           V: dace.float32[N, d], O: dace.float32[N, d]):
4
5      for ti in dace.map[0:N:Ti]:
6
7          m = np.full([Ti], -np.inf, Q.dtype)
8          l = np.zeros([Ti], Q.dtype)
9          S = np.empty([Ti, Tj], Q.dtype)
10         Oi = np.zeros([Ti, d], Q.dtype)
11
12         for tj in range(0, N, Tj):
13
14             S[:] = Q[ti:ti+Ti, :] @ K[tj:tj+Tj, :].T
15
16             max_row = np.max(S, axis=1)
17             m_new = np.maximum(m, max_row)
18             p_tilde = np.exp(S - m_new[:, np.newaxis])
19             sum_row = np.sum(p_tilde, axis=1)
20             l_tmp = l * np.exp(m - m_new)
21             l_new = l_tmp + sum_row
22             Oi[:] = (Oi * l_tmp[:, np.newaxis] + p_tilde @ V[tj:tj+Tj, :])
23                     / l_new[:, np.newaxis]
24             m[:] = m_new
25             l[:] = l_new
26
27         O[ti:ti+Ti, :] = Oi
```

Figure 17: DaCe data-centric Python implementation of Flash Attention.

### 9.6. Flash Attention with APL

The APL code is a fairly direct translation of the FlashAttention algorithm, but cannot control memory allocation and, therefore, cannot exploit the inner loops working on shared memory objects. Hence, the implementation is closer to Algorithm 6. Input values are converted to 3-dimensional arrays to facilitate blocking before entering the main loop. While the other languages can express single precision floating point computation, existing APL implementations only support double precision floating points or greater.

### 9.7. Flash Attention with DaCe

For CPU, we use the Python code shown in Fig. 17 with auto-optimization heuristics. DaCe generates OpenBLAS calls for the matrix multiplications in lines 16 and 25). We explicitly set the number of OpenBLAS threads to one, resulting in a very similar implementation to the CPU baseline code. For GPU, we use the simpler implementation shown in Algorithm 6.

### 9.8. Summary

The Flash Attention performance results for the five languages on both multicore CPU and GPU are summarised in Table 8. We follow the baseline performance measurement and report compute rate (Gflops).

Table 8: FlashAttention compute rate in Gflops (higher is better) for different $N$ and $d$.

|  | $(d, N) = (64, 16384)$ | | | $(d, N) = (64, 32768)$ | | |
|---|---|---|---|---|---|---|
|  | CPU | | GPU | CPU | | GPU |
|  | 1C | 32C |  | 1C | 32C |  |
| Baseline | 78 | 2030 | 5669 | 77 | 2238 | 6585 |
| Accelerate | 73 | 96 | 235 | 150 | 399 | 571 |
| APL | 15 | — | 266 | 15 | — | 295 |
| DaCe | 72 | 1666 | 2073 | 73 | 1897 | 1303 |
| Futhark | 4 | 84 | 3668 | 4 | 84 | 3704 |
| SaC | 26 | 666 | - | 26 | 714 | - |

|  | $(d, N) = (128, 8192)$ | | | $(d, N) = (128, 16384)$ | | |
|---|---|---|---|---|---|---|
|  | CPU | | GPU | CPU | | GPU |
|  | 1C | 32C |  | 1C | 32C |  |
| Baseline | 81 | 2222 | 5416 | 80 | 2375 | 6573 |
| Accelerate | 18 | 48 | 111 | 36 | 182 | 228 |
| APL | 25 | — | 363 | 25 | — | 574 |
| DaCe | 88 | 1632 | 2997 | 88 | 2136 | 3669 |
| Futhark | 3 | 73 | 4589 | 2 | 67 | 4404 |
| SaC | 27 | 608 | - | 26 | 702 | - |

*GPU performance.* All languages implement the less efficient Custom Attention (Algorithm 6), and hence cannot match the baseline performance. Futhark and DaCe are the closest, reaching (65 %, 56 %, 85 % and 67 %) and (37 %, 20 %, 55 % and 56 %) of the baseline performance on the four datasets respectively. APL and Accelerate lack locality optimizations, which restricts the percentage of baseline performance to (4.7 %, 4.5 %, 6.7 % and 8.7 %) and (4.1 %, 8.7 %, 2.5 % and 3.5 %), respectively. Compiler shortcomings prevent a SaC implementation.

*Multicore performance.* DaCe reaches a good fraction of the baseline performance 82 %, 85 %, 73 % and 90 %, followed by SaC at 33 %, 32 %, 22 % and 30 %. Accelerate and Futhark suffer from a lack of locality optimizations, reaching only (4.7 %, 18 %, 2.2 % and 7.7 %) and (4.1 %, 3.8 %, 3.3 % and 2.8 %) of the baseline performance. We did not benchmark APL using manual task parallelism for multicore CPUs, and the implicit multi-threading of the interpreter did not significantly impact results, so we include only a single result for each CPU result for APL.

*Expression.* While Custom Attention is elegantly implemented by all languages, with code that closely matches Algorithm 6, we cannot meaningfully compare most languages with the baseline because it implements the more efficient Flash Attention. The exception is DaCe's Multicore implementation of Flash Attention (Fig. 17) that is very close to the algorithmic specification.

## 10. Benchmarking Summary and Analysis

Measuring the performance of the five languages on such a challenging set of benchmarks on both a multicore and a GPU exposes their design and implementation strengths and weaknesses. Broadly speaking, a language delivers good performance on a target architecture if it has the right constructs for specifying the benchmark, paired with the right optimizations to generate efficient code for the target architecture. Conversely, performance is lacking if the language lacks the required constructs, optimizations, or, indeed, an implementation for the target architecture.

We find that *functional array languages are expressive: they can represent a variety of array applications both concisely, and preserving a close relationship to the high-level specification.* Detailed evidence for this claim is provided in the summary of each benchmark, i.e., in Sections 6.8, 7.8, 8.8 and 9.8, and some specific examples follow. SaC's MG implementation closely follows the mathematical formulation presented in Section 7.1 and Algorithm 1. Notably, it defines one two-line generic stencil kernel (Fig. 12) that is instantiated in different contexts (i.e., to perform the computations denoted by $A$, $P$, $Q$, $S$). In contrast, the CUDA/Fortran baselines specialize each stencil kernel into different code ultimately resulting in a code base that our benchmark implementation teams found extremely hard to understand. Futhark builds on SaC's implementation and demonstrates that the baseline's optimizations can be expressed at a high level (i.e., applied to the generic stencil kernel), as evidenced by the competitive GPU performance. DaCe's base $N$-body implementation consists of about ten lines of Python code, similar to the five equations describing the problem. In contrast, the baseline consists of three code versions (one for CPU and two for GPUs) that are optimized in very different ways, e.g., they utilize different levels of parallelism and different array layouts (SoA *vs.* AoS). For Quickhull, Accelerate and Futhark present reasonably straightforward manually-flattened implementations that build on their rich set of second-order array operators and guarantee deterministic-by-construction parallelism. Ultimately, all arguments related to "elegant expression" are subjective, and we encourage readers to inspect our code repository and draw their own conclusions.

As an approximate measure of code size and development effort we use Source Lines Of Code (SLOC), a crude but widely accepted measure [8]. SLOC counts the lines of code omitting blank and comment lines. Table 4 compares the benchmark sizes. The precise numbers must be taken with a grain of salt, as they do not account for differences in personal style and tooling—for example, some languages use external tooling for timing and input processing, while others implement it for each benchmark. The major trends, however, are not obscured by such issues. A major difference is that the functional array programmer writes a single program that is compiled for multiple targets, where there are separate CPU and GPU baselines. Hence the bottom two rows of the table show that *at 7633 SLOC the total baseline codebase is much larger, at least 10× than any of the functional array codebases (683 SLOC).* The functional code requires fewer lines of code than the baselines in every case except for

|  | Accelerate | APL | DaCe (CPU) | DaCe (GPU) | Futhark | SaC | Baseline (CPU) | Baseline (GPU) |
|---|---|---|---|---|---|---|---|---|
| $N$-body | 113 | 25 | 76 | 53 | 46 | 61 | 164 | 411 |
| MG | 137 | 27 | 255 | | 151 | 136 | — | 4662 |
| Quickhull | 254 | 26 | — | — | 161 | 203 | 208 | — |
| FlashAttention | 179 | 21 | 31 | 26 | 90 | 79 | 1280 | 908 |
| Total Codebase | 683 | 99 | — | — | 448 | 479 | 1652 | 5981 |
| | | | | | | | | 7633 |

Table 9: Benchmark sizes in Source Lines of Code (SLOC).

Accelerate Quickhull on CPU. *The functional codebases are at least $2\times$ smaller than the CPU baseline codebase (683 vs 1652 SLOC) and much smaller, at least $8\times$, than the GPU baseline codebase (683 vs 5981 SLOC).* Of the functional languages APL is the most concise (99 SLOC in total), while Accelerate is the most verbose (683 SLOC in total).

Regarding GPU and multicore performance, there are a total of 36 benchmark instances with a baseline. In 30 % of the instances (11 instances), at least one language matches or outperforms the baseline. In 70 % of the instances (25 instances), at least one language achieves more than 80 % of the baseline performance. In a further 25 % of the instances (9 instances), at least one language achieves more than 50 % of the baseline performance. In only 6 % of instances (2 instances), no language achieves more than 50 % of the baseline performance. These instances are Quickhull solving the Circle and Rectangle datasets on a 32-core CPU, and the reasons are elaborated in Sections 8.2 and 8.8. *The key conclusion from these results is that they demonstrate that mature functional array languages have the potential to deliver performance competitive with the best available conventional techniques.*

The remainder of the section briefly summarizes and analyses the performance of each language on the set of benchmarks and target architectures.

*10.1. Futhark Summary and Analysis*

Futhark's GPU performance is competitive with the baselines for all benchmarks other than Flash Attention. That is, Futhark sometimes outperforms the baseline, and the speedup is always more than $0.97\times$ the baseline. A key factor for achieving good performance is the compiler's ability to exploit nested parallelism at the application level by mapping it to different levels of the hardware hierarchy. The key optimization is *incremental flattening* ($\mathcal{IF}$), a multi-version compilation technique (Section 4.1). For $N$-body, $\mathcal{IF}$ generates two versions of the code, one that utilizes only the outermost level of parallelism for small numbers of bodies, and one that parallelizes both outer and inner levels of parallelism for large numbers of bodies. For MultiGrid andFlash Attention, $\mathcal{IF}$ generates code versions that map the innermost level of parallelism to the CUDA thread block level, enabling reuse from fast memory.[10] Finally, Quickhull benefits from efficient GPU implementations of flat-parallel constructs such as scan and reduce-by-index, and Flash Attention from block-and-register tiling optimization of batched matrix multiplications. All benchmarks benefit from fusion [55], which is enabled by IR constructs that allow, e.g., to also return the mapped part of a map-reduce composition [56], if this is needed elsewhere.

Futhark's CPU backend uses a different compilation pipeline that has received less attention than the GPU. Notably, it lacks locality optimizations, resulting in abysmal performance for Flash Attention, and it employs a runtime system that aims to dynamically adjust the levels of parallelism that are mapped to the hardware, but which introduces significant overhead for *N*-body and MG.

*10.2. Accelerate Summary and Analysis*

All the Accelerate benchmark programs are straightforward implementations of the algorithms, without any attempt to tune them for performance. Furthermore, none of the benchmarks utilise bindings to high-performance libraries like BLAS. Accelerate's foreign function interface makes it easy to use such high performance libraries. The results reveal that Accelerate has a relatively high performance overhead, although for large computations the overhead is amortized. The Accelerate compiler pipeline is currently being completely rewritten. As a consequence, the back-ends haven't received major updates for several years. In particular, the GPU back-end has not been optimized for more modern GPU architectures, and the effect can be seen in the benchmarks.

*10.3. SaC Summary and Analysis*

SaC's CPU performance is competitive with the baseline in *N*-body. For both Multigrid and Flash Attention the source code of SaC is much closer to the mathematical specification than the baseline. This clarity costs a factor $1.2 - 3.9$ performance. The key optimization is *with-loop folding*, which improves the temporal locality by composing functions element-wise.

The SaC GPU backend is not as well developed and the benchmark performance does not come close to the baseline.

*10.4. APL Summary and Analysis*

APL is by far the most concise language, and the computational core of each benchmark is implemented using between 10–14 lines of code. The amount of native data type, functional, and structural flexibility in the APL language exceeds that of the other languages in this comparison. It also has the richest native set of primitive operations over its core data structure, the array.

---

[10] As such versions may run out of resources (e.g., fast memory) for datasets with a large innermost parallel dimension, they require multi-version compilation.

However, both the Dyalog CPU interpreter and Co-dfns GPU compiler lack the full set of optimizations to extract the maximum performance from these benchmarks. In both implementations, the primary factor limiting performance is the inability to eliminate memory access costs associated with intermediate and temporary variables. For instance, the current runtime requires the use of double-precision floating points, which harms performance on the Flash Attention benchmark. Likewise, because the GPU runtime depends on runtime level JIT compilation provided by external libraries [35], there is some variability and lack of predictability in the level of fusion and the kind of laziness that will manifest. This causes some benchmarks to perform worse than strictly necessary for the given APL code. For the QuickHull benchmark, runtime JIT does not optimize the implementation of key-based reduction, which must be implemented using other primitives, since APL does not presently support key-wise reduction. This greatly increases both the number of kernels and the main memory throughput required. APL's dynamic typing makes type inference more difficult, limiting the ability to statically specialize the output to known data sizes. This requires runtime verification of some data types, array shapes, and data ranges to implement the primitives, further increasing overheads.

While it is possible to manually tile or tweak the APL code to encourage the runtime JIT to produce more optimal code, doing so is unidiomatic. The benchmarks are expressed as idiomatic APL without such manual optimization.

The Co-dfns GPU compiler is relatively new, and is the least optimizing compiler among those evaluated here. There is very little optimization of locality between kernels or between various computations. So when a computation maps to a specific library function it is performant, but combining unfused primitives typically requires writing to GPU main memory, limiting performance. Extending Co-dfns with additional optimizations would permit more efficient kernel generation and resolve many of these issues.

While the lack of optimizations limits APL performance, the Co-dfns compiler consistently delivers more than a $10 \times$ performance improvement over the Dyalog CPU interpreter. This shows that APL is eminently suitable for GPU execution, and the ease with which it can be translated for the GPU, even without optimizations and specialized compilation pipelines. This is rare, since most languages rely on specific compiler optimizations to achieve good GPU performance.

*10.5. DaCe Summary and Analysis*

DaCe's greatest asset is undoubtedly its extensive optimization arsenal. The automatic tools frequently lead to acceptable performance, while experimenting with different parallel schedules and data layouts is often relatively easy. At the same time experienced users can further improve performance by manually applying transformation workflows, by editing the graph representation with visual tools, or even by embedding scheduling decisions in the high-level code. These capabilities can deliver high performance, as shown in $N$-body, where DaCe matches the baseline on CPU and surpasses it on GPU by 35 % and 23 % on the medium and large datasets, respectively. However, taking advantage of such tools requires both a level of expertise and investing some time.

Another of DaCe's strengths is its Library Node system, which simplifies integration with optimized libraries like vendor-provided BLAS/LAPACK implementations. This leads to high performance in benchmarks such as Flash Attention, where DaCe achieves 80 % to 90 % of the baseline's performance on CPU.

A significant drawback for some algorithms, however, is the lack of support for recursion. For example the lack of recursion constrains DaCe to an imperative implementation of MG, and makes the generation of a graph representation a difficult task in Quickhull.

## 11. Related Work

There are a multitude of parallel programming paradigms and a range of target architectures. We do not replicate the detailed discussions [4, 1, 15, 6, 7] of how each of the functional array languages covered here is situated in the parallelism space. Rather, we focus on describing the subspace in the parallel paradigm design space occupied by functional array languages.

*Low* vs *High Level, Task* vs *Data Parallelism.* Some parallel paradigms are *low-level* and others are *high-level*. In low-level paradigms, the parallel coordination is specified in great detail, giving the programmer a high degree of control, but also a significant coding challenge. For example, many MPI [57] programs use individual message `send`s that must be exactly matched with message `receive`s. In contrast, high-level paradigms have powerful coordination constructs, and this is the approach adopted by functional array languages. The coordination constructs are typically a set of parallel combinators, that is, polymorphic higher-order functions with implicit parallel semantics. A simple example is a parallel `map` that applies a function to every element of an array in parallel.

Some parallel paradigms use *task* parallelism, where tasks are spawned and collaborate to achieve the required computation. Examples include MPI and some parallel functional languages like Glasgow parallel Haskell [58]. Other paradigms use *data* parallelism, where the parallelism is determined by the data structures being manipulated. Functional array languages take the latter approach. More precisely, they provide *bulk* data parallelism where the combinators operate over an entire multidimensional array.

Some parallel paradigms are implemented as a library, like NumPy [59]or MPI [57], while others are provided as a programming language, or language extension, e.g., SaC extends C. Functional array languages are typically languages or DSLs, and rely on sophisticated compilers to generate performant code. The following discussion focuses primarily on parallel languages.

*Parallel Hardware.* Some parallel paradigms target shared-memory architectures, some distributed architectures, and others both. Parallel languages that can target distributed memory architectures, like Chapel [60], Charm++ [61], HPF [62], and X10 [63] scale across multiple hosts and may execute on a cluster or in a cloud. In contrast, shared-memory paradigms like OpenMP [64] utilize the resources of a single host, possibly in conjunction with accelerators. Most of the languages discussed here are primarily shared-memory parallel languages, although APL, DaCe, and SaC have the capability to target clusters (Table 3).

Parallel architectures are increasingly heterogeneous, providing accelerators alongside multicores. Common accelerators are GPUs, FPGAs, or AI accelerators. All of the functional array languages discussed here target GPUs in addition to multicores, and DaCe can target FPGAs (Table 3). Some widely-used paradigms targeting accelerators are low-level, like CUDA [65] for GPUs, and OpenCL [66] for FPGAs and GPUs, while others offer replacements of mainstream libraries, e.g., cuNumeric [67] and CuPy [68] for NumPy. A range of other languages are similar to functional array languages in aiming to provide high-level accelerator programming, e.g., Chapel [60] and X10 [63].

*Imperative Approaches.* A large body of work is dedicated to libraries, runtime systems, and compilation techniques aimed at supporting heterogeneous/portable (HPC) programming within mainstream imperative languages. Examples of library extensions of C++ include Kokkos [69] and RAJA [70], and runtime systems include HPVM [71] and Legion [72]. Polyhedral analyses [73], integrated into optimizing compilers such as Pluto [74], PPCG [75], Polly [76], Tiramisu [77] have demonstrated that at least *affine* programs can be efficiently mapped to multi-core and GPU hardware. The optimization spaces for polyhedral analysis and the rewriting in functional array languages both overlap and are complementarity. For example, map fusion/fission are analogous to loop fusion/distribution. On the other hand, low-level transformations requiring dependence-analysis on arrays, such as loop skewing, are not expressible as functional rewrites. Conversely, polyhedral analysis is limited to affine code[11]—consisting of loop nests in which all control structures and array indices are affine expressions in terms of the surrounding-loop indices—hence dynamic control flow and scatter/gather operations cannot be analyzed. Furthermore, polyhedral analysis lacks support for second-order parallel primitives such as (segmented) scan[12] (a.k.a., prefix sum). Hence computations that are dominated by scans, like the Quickhull benchmark, fall outside the polyhedral scope.

*DSLs and Scheduling Languages.* Conventional compilers automatically generate low-level code using a set of optimization heuristics. For performance-critical applications, the heuristics often fall short of extracting optimal performance because (i) they cover a small part of a huge optimization space, and (ii) they typically generate one optimized code version, that may not be ideal for all datasets of interest. This has motivated the emergence of scheduling languages (surveyed in [84]) that specify optimization recipes as a combination of code transformations both as a way of allowing the user to take an active hand in the optimization process, and/or to enable systematic exploration of the optimization space. Early approaches, such as the X language [85], CHiLL [86], URUK [87], expose polyhedral transformation that are applied to scientific code written in mainstream languages such as C.

---

[11]A significant body of work was aimed at relaxing the affine constraints, by combining static with more dynamic instances of dependence analyses on arrays [78, 79, 80, 81, 82, 83].

[12]A sequential implementation of scan is affine code, but the transformation to a parallel scan requires an algorithmic change that is not expressible in terms of polyhedral transformations.

More recent scheduling languages are driven by the observation that specialization of the language and compiler repertoire paves the way to performance. They commonly employ a simple (often functional) specification language, that is accessible to domain experts, and a separate optimization recipe written by compiler experts. Classic examples include (i) image-processing DSLs like Halide [88], PolyMage [89], Darkroom [90], (ii) tensor algebra DSLs for dense computations, such as Baracuda [91], DISTAL [92], and for sparse computations, such as CHiLL-I/E [93, 94], TACO [95], Mosaic [96], and (iii) machine learning DSLs such as Triton [97], TVM [98] and SWIRL [99]. The functional languages studied in this paper are more general-purpose than these DSLs as they aim to parallelize any array-based computation. Among them, only DaCe has support for user-defined schedules, while Futhark uses a specialized compiler exploration strategy that generates multiple semantically equivalent code versions that are discriminated by autotuning, as described in Section 4.1.

*Other Functional Array Languages.* This paper focuses on a representative set of functional array languages, but there are others that we have not covered. For example, Dex [100] is very similar to the languages covered here. Other examples are the LIFT [101], RISE [102] and MDH[103] languages that are similar to, albeit more restricted than the languages discussed in this paper, but may support user-guided exploration of the optimization space [104]. Unlike the languages in this paper, which all restrict higher-order functions, Erik Holk's Harlan language [105] supports first class procedures natively.

## 12. Discussion and Conclusion

Functional array languages offer the enticing prospect of the high-level specification of correct-by-construction parallel array programs that generate performant and portable code. Hence, they are attracting research interest, and are emerging as a class of high-performance parallel languages. This paper compares the design, implementation, and performance of five functional array languages, namely Futhark, Accelerate, SaC, APL, and DaCe, and makes the following research contributions.

We illustrate the programming paradigms of the five functional array languages using a simple *N*-body benchmark (Section 2). The five *N*-body implementations are both more concise than the CUDA and OpenMP baseline implementations and, we argue, far closer to the mathematical specification.

We make a systematic comparison of the designs of the five functional array languages (Section 3). All of the languages support rectilinear arrays, several support rank polymorphism, and user-defined element types, but only APL supports jagged arrays and heterogeneous arrays (Table 2). All our languages except APL use static types, most support parametric polymorphism, and only Futhark supports dependent types (Table 1). Although most languages offer at least partial determinism, the languages otherwise offer a wide variety of parallel paradigms: parallelism may be implicit or explicit, may be nested, and task parallelism may be available alongside the data parallelism (Table 3).

We outline the implementation of the five functional array languages, and make a systematic comparison (Section 4). There are a variety of implementation techniques: the Futhark and SaC compilers, the Accelerate DSL, the DaCe framework, and the APL interpreter and compiler. All implementations target multicore CPUs and GPUs, and several languages target other platforms. Although all implementations fuse computations where appropriate, implementations otherwise support different sets of optimizations (Table 4).

We demonstrate the expressiveness of functional array programming using four challenging benchmarks that represent a range of application domains, parallel computational models, and exploit different optimizations (Sections 6 to 9). The benchmark code is developed iteratively by applying a sequence of optimizations to an initial correct implementation.

We compare the codebase sizes, and hence approximate the programming effort, of the benchmarks in the functional array languages and in the OpenMP and CUDA baselines. We use Source Lines Of Code (SLOC) as a crude but widely accepted measure [8] (Table 4). A major difference is that the functional array programmer writes a single program that is compiled for multiple targets, where there are separate CPU and GPU baselines. Hence *the total baseline codebase (7633 SLOC) is much larger, at least 10× than any of the functional array codebases (683 SLOC)*. The functional code requires fewer lines of code than the baselines in all but one case. *The functional codebases are at least 2× smaller than the CPU baseline codebase and much smaller, at least 8×, than the GPU baseline codebase.* Of the functional languages APL is the most concise, with Accelerate and SaC being the most verbose.

One reason for their conciseness is that the functional array code omits architectural aspects like the memory hierarchy and how the algorithm is mapped to hardware. This enables the development of code without intimate knowledge of the parallel hardware. It also allows the implementation to generate both multicore and GPU programs from a single source. Thus, *functional array code should be more easily ported to, and optimized for, new parallel architectures.* The challenge of porting functional array programs to some new hardware, perhaps neuromorphic computing, is that of writing a performant implementation, like a compiler backend, for the new hardware. The evidence for this claim in Sections 6 to 9 is partial as only a few languages, notably Dace and Futhark, consistently achieve good performance on both multicore and GPU. We argue that the primary reason that other languages don't deliver good performance on both architectures is language engineering: the research teams have not yet had the resources to engineer a well performing implementation.

We summarize and analyze the multicore CPU and GPU performance of the languages on 39 instances of the four open-source benchmarks[13] to explore why each language performs well, or poorly, on each benchmark and architecture (Section 10). We find that in 30 % of the instances, at least one language matches or outperforms the baseline; that in 70 % of the instances at least one language achieves more than 80 % of the baseline performance; and in 94 % of the instances at least one language achieves more than 50 % of the baseline performance. In only 6 % of instances, no language achieves more than 50 % of the baseline performance. These results are impressive considering how new the language implementations are, and *we argue that the results demonstrate that mature functional array languages have the potential to deliver performance competitive with the best available conventional techniques.* We look forward to seeing the developments in the area in the coming years.

*Funding Sources*

This work was funded in part by the following research grants. Novo Nordisk Foundation (grant reference number NNF24OC0090447). Swiss National Science Foundation (SNSF), grant agreement "Quantum Transport Simulations at the Exascale and Beyond (QuaTrEx)" no. 209358. UK EPSRC STARDUST grant number EP/T014628.

---

[13] https://github.com/diku-dk/CFAL-bench/